\documentclass[aps,pre,reprint,superscriptaddress]{revtex4-2}
\usepackage{eurosym}
\usepackage{graphicx}
\usepackage{amsmath}
\usepackage{latexsym}
\usepackage{amsthm, amscd, amsfonts, amssymb}
\usepackage{multirow}
\usepackage{makecell}
\usepackage{dcolumn}
\usepackage{tabularx}
\usepackage{xcolor}
\usepackage{etoolbox}
\usepackage{threeparttable}

\begin{document}

\title{Multi-pole solitons and breathers with spatially periodic modulation
induced by the helicoidal spin-orbit coupling}

\author{Cui-Cui Ding}
\affiliation{Research Group of Nonlinear Optical Science and Quantum Technology, \\ School of Microelectronics, Wuhan Textile University, Wuhan 430200, China}
\affiliation{Research Center of Nonlinear Science, School of Mathematical and Physical Sciences,\\ Wuhan Textile University, Wuhan 430200, China}
\author{Qin Zhou}
\email{qinzhou@whu.edu.cn}
\affiliation{Research Group of Nonlinear Optical Science and Quantum Technology, \\ School of Microelectronics, Wuhan Textile University, Wuhan 430200, China}
\affiliation{Research Center of Nonlinear Science, School of Mathematical and Physical Sciences,\\ Wuhan Textile University, Wuhan 430200, China}
\affiliation{State Key Laboratory of New Textile Materials and Advanced Processing Technologies,\\ Wuhan Textile University, Wuhan 430200, China}
\author{B. A. Malomed}
\affiliation{Department of Physical Electronics, School of Electrical Engineering,
	Faculty of Engineering, and the Center for Light-Matter University, Tel Aviv
	University, Tel Aviv, Israel}
\affiliation{Instituto de Alta Investigaci\'{o}n, Universidad de Tarapac\'{a},
Casilla 7D, Arica, Chile}

\begin{abstract}
We report analytical solutions for diverse multi-pole (MP) soliton and
breather states in spatially inhomogeneous binary Bose-Einstein condensates
(BECs) with the helicoidally shaped spin-orbit coupling (SOC), including MP
stripe solitons on zero background, MP beating stripe solitons on a nonzero
plane-wave background, as well as MP beating stripe solitons and MP
breathers on periodic backgrounds. The results indicate that modulation
effects produced by the helicoidal SOC not only induce stripe patterns in MP
solitons, but also generate the spatially-periodic background for the MP
beating stripe solitons and breathers. An asymptotic analysis reveals curved
trajectories with a logarithmically increasing soliton/breather separation
for these MP excitations, fundamentally distinguishing them from periodic
trajectories of bound-state solitons/breathers or straight trajectories of
conventional multi-soliton/breather sets. With complex periodic structures
in individual components, the total density distribution is nonperiodic, due
to their configurations which are out-of-phase with respect to the two
components. We further examine several degenerate structures of MP solitons
and breathers under varying SOC and spectral parameters. Numerical
simulations validate the analytical results and demonstrate stability of
these MP excitations. These findings may facilitate deeper understanding of
soliton/breather interactions beyond conventional multi-soliton systems and
bound-state complexes in SOC BEC.
\end{abstract}

\maketitle

\section{Introduction}


Multi-pole (MP) solitons, which are characterized by curved trajectories of
the poles and logarithmically increasing separation between them, have
attracted much interest in the form of multi-soliton systems~\cite%
{Zakharov1972,SY,Zakharov,Olmedilla1987,Gagnon1994,DP,Yang,Schiebold2017} --
in particular, for investigating soliton-interaction dynamics~\cite%
{Zakharov2013,Kengne2023} and breathers, built as a nonlinear superposition
of two or several solitons with a common center~\cite{SY}. A particularly
fascinating aspect of the soliton physics is their particle-like interaction
dynamics: solitons may exhibit attractive collisions, repulsive scattering,
or mutual annihilation, in some cases~\cite%
{Zhang2024,Poy2022,Krolikowski1998}. As an application, the soliton-like bubble excites the cavitation to launch the microbot~\cite{Wang2025a}. Parallel to experimental investigations~%
\cite{Wang2025}, theoretical studies extensively explored multi-soliton
interactions through constructed multisoliton solutions of nonlinear
evolution equations~\cite{Ablowitz2023,Abbagari2025}. These solutions,
corresponding to distinct discrete complex spectral parameters (which
represent simple poles, in terms of the inverse-scattering transform \cite%
{Zakharov1972,Zakharov}) describe elastic or inelastic interactions between
solitons \cite{Ablowitz1992,Zakharov1972a,Xin2022}.

Growing interest is drawn to two special forms of the multisoliton
solutions. One is the class of bound-state solitons (e.g,, \textquotedblleft
soliton molecules" ) which are formed when spectral parameters share
identical real parts but distinct imaginary parts, exhibiting periodic
attractive/repulsive forces~\cite{Qin2019,Cui2024}. The other
special class represents MP solitons, which emerge when spectral parameters
coincide (with the eigenvalue degeneracy higher than one), creating
nonperiodic weakly bound states with strong near-field interactions \cite%
{Zakharov1972,Olmedilla1987,Gagnon1994,Schiebold2017}. Unlike conventional
solitons (with constant velocities) or bound states featuring periodic
oscillations, MP solitons follow curved trajectories with time-dependent
velocities and logarithmically increasing separations, indicating sustained
attraction~\cite{Li2020,Li2020a,Xu2019a}.

Double-pole (DP) solitons were first proposed by Zakharov and Shabat \cite%
{Zakharov1972} and Satsuma and Yajima \cite{SY}, as degenerate two-soliton
solutions of the nonlinear Schr\"{o}dinger equation (NLSE)~. Subsequent
studies established rigorous asymptotic descriptions of general MP solitons,
using the operator-theoretic approach~\cite{Schiebold2017}. MP solitons have
been extensively investigated in integrable systems, including the modified
Korteweg - de Vries and sine-Gordon equations~\cite{Poppe1983,Wadati1981},
and recently extended to nonlocal~\cite{Wang2024} and multidimensional \cite%
{Zhao2020} systems. DP soliton-like solutions with a logarithmically growing
separation between the two constituents exist even in nonintegrable versions
of NLSEs~\cite{Nguyen2019}. Multicomponent systems also host MP solitons
with intriguing dynamics~\cite{Rao2020,Rao2021}, studied across
hydrodynamics, Bose-Einstein condensates (BECs), and nonlinear optics~\cite%
{Gagnon1994,Gordon1983,Karlsson1996}. In particular, in optical fibers, MP
solitons model interactions of chirped in-phase pulses with identical
amplitudes and group velocities \cite{Gagnon1994,Gordon1983}. Breathers,
which represent another class of localized excitations, characterize
evolving perturbations on finite backgrounds~\cite{Xu2019}. Beyond the MP
solitons, MP breathers have gained interest too, particularly regarding
their connection to rogue-wave generation mechanisms~\cite%
{Kedziora2012,Liu2017}.

A milestone achievement in ultracold atomic physics was the realization of
momentum-dependent artificial gauge potentials through Raman-laser-induced
coupling, enabling electrically neutral cold atoms to exhibit spin-orbit
coupling (SOC) phenomena similar to charged particles in electromagnetic
fields~\cite{Lin2011,Zhai2015,Zhang2016}. Building on this foundation,
research into artificial vector gauge potentials has become feasible~\cite%
{Dalibard2011,Ruseckas2005}. When spatial modulations satisfy specific
symmetries, the constraint preventing free propagation of nonlinear
localized excitations due to broken translational invariance is overcome~%
\cite{Kartashov2017,Kartashov2019,Fang2024}. In particular, Ref. \cite%
{Kartashov2017} has introduced helicoidal SOC via spatially inhomogeneous
gauge potentials, and found that freely moving solitons can stably propagate
in the corresponding spatially inhomogeneous BECs with helicoidal SOC.
A kind of kink-like breathers, characterized by the difference between
the background densities on their two sides (the kink's height), have been reported
in the BEC with helicoidal SOC~\cite{Yang2022}.

In this work, we investigate the control effects of helicoidal SOC on
various MP solitons/breathers in BECs, particularly focusing on stripe
states induced by SOC and beating structures arising from the dark/bright
soliton superposition. First, through the gauge transformation applied to
the Manakov system, we construct general MP soliton/breather solutions.
Subsequently, a comprehensive analysis of these solutions are conducted for
the cases of zero, plane-wave, and spatially periodic backgrounds,
emphasizing the impact of helicoidal SOC on their dynamical properties.
Finally, numerical simulations validate the solutions' robustness and
stability against perturbations.

The paper is structured as follows. Section II introduces the helicoidal
SOC-BEC model and its general MP soliton/breather solutions. Vector
bright-bright double- and triple-pole stripe solitons are investigated in
Section III. Section IV examines vector MP beating stripe solitons on the
nonzero plane-wave background, and their degenerate forms. Section V
explores the dynamics of MP beating stripe solitons and MP breathers on
periodic backgrounds, in wavenumber-matched and mismatched regimes.
Numerical simulations for MP solitons/breathers are presented in Section VI,
The paper is concluded in Section VII.

\section{The model and general MP (multi-pole) soliton and breather solutions}

The one-dimensional BEC with helicoidal SOC is governed, in the mean-field
approximation, by the system of coupled Gross-Pitaevskii equations~\cite%
{Kartashov2017}:
\begin{equation}
i\frac{\partial \mathbf{\Psi }}{\partial t}=\frac{1}{2}Q^{2}(x)\mathbf{\Psi }%
-s(\mathbf{\Psi }^{\dag }\mathbf{\Psi })\mathbf{\Psi },
\label{helicoidal SOC}
\end{equation}%
where $\mathbf{\Psi }=(\Psi _{1},\Psi _{2})^{T}$ is the spinor wave function
with interatomic attractive or repulsive interactions, defined, by $s=+1$
and $-1$, respectively. The system incorporates experimentally tunable SOC
strength $\alpha $, with the helicoidal structure defined by means of the
generalized momentum operator \cite{Jimenez2015,Luo2016} $Q(x)=-i\partial
/\partial x+\alpha \boldsymbol{\sigma }\cdot \mathbf{n}(x)$, with the
spatial modulation represented by $\mathbf{n}(x)=(\cos (2\kappa x),\sin
(2\kappa x),0)$. %\begin{equation}
%Q(x)=-i\partial /\partial x+\alpha \boldsymbol{\sigma }\cdot \mathbf{n}(x)
%\label{Qx}
%\end{equation}
%with the spatial modulation defined as
%\begin{equation}
%\mathbf{n}(x)=(\cos (2\kappa x),\sin (2\kappa x),0).
%\label{spatial modulation}
%\end{equation}
Here, $\kappa >0$ and $\kappa <0$ correspond to the right- and left-handed helicity, respectively~\cite{Samsonov2004,Burt2004}, and $%
\boldsymbol{\sigma }=(\sigma _{1},\sigma _{2},\sigma _{3})$ is the vector of
the Pauli matrices. Eq.~(\ref{helicoidal SOC}) reduces to the
homogeneous Rashba-Dresselhaus SOC when $\kappa =0$~\cite{Dalibard2011}, and
to the canonical Manakov system when $\alpha =0$~\cite{Manakov1974}.

To construct MP solitons and breather solutions of Eq.~(\ref{helicoidal SOC}%
), we first make a spatially-dependent substitution,
\begin{equation}
\begin{aligned} \label{trans1} \mathbf{\Psi
}=\mathcal{G}\mathbf{u}=\begin{pmatrix} \nu _{+}e^{-i(k_{\text{m}}+\kappa
)x} & \nu _{-}e^{i(k_{\text{m}}-\kappa )x}\\ \nu
_{-}e^{-i(k_{\text{m}}-\kappa )x} & -\nu _{+}e^{i(k_{\text{m}}+\kappa )x}
\end{pmatrix}\mathbf{u}, \end{aligned}
\end{equation}%
to transform this equation into the integrable Manakov system~\cite%
{Manakov1974}
\begin{equation}
i\mathbf{u}_{t}+\frac{1}{2}\mathbf{u}_{xx}+s(\mathbf{u}^{\dag }\mathbf{u})%
\mathbf{u}=0,~~\mathbf{u}=(u_{1},u_{2})^{T},  \label{Manakov}
\end{equation}%
where
\begin{equation}
\begin{aligned} \nu _{+}=&\text{sgn}(\alpha )\sqrt{\left( k_{\text{m}}-
\kappa \right) /\left( 2k_{\text{m}}\right) }, \\ \nu
_{-}=&\sqrt{\left(k_{\text{m}}+ \kappa \right) /\left(
2k_{\text{m}}\right)}, \end{aligned}  \label{nu}
\end{equation}%
and $k_{\text{m}}=\sqrt{\alpha ^{2}+\kappa ^{2}}$ is the effective momentum
of the lowest-energy states. The $x$-dependence of the transformation matrix
$\mathcal{G}$ in transformation (\ref{trans1}) is the origin of the striped
structures in MP solitons and spatiotemporal periodic background in MP
breathers which are considered below. A more general form of
transformation (\ref{trans1}) was proposed and employed to investigate a more general
spatially inhomogeneous SOC-BEC model in works \cite{Kartashov2014,Kartashov2019}.

Starting from zero/plane-wave initial states, substitution~(\ref{trans1}),
coupled to the Manakov-type generalized Darboux transformation~\cite{Guo2012}%
, generates the MP soliton/breather solutions for system~(\ref{helicoidal
SOC}) through:
\begin{equation}
\begin{aligned} \label{multi-pole} \begin{pmatrix} \Psi_1 \\ \Psi_2
\end{pmatrix}=\mathcal{G}\begin{pmatrix} u_{10}+
\frac{2}{|\mathbb{W}|}\begin{vmatrix} \mathbb{W} & \mathbb{Y}_1^{\dag} \\
\mathbb{Y}_2 & 0 \end{vmatrix} \\[4mm] u_{20}+
\frac{2}{|\mathbb{W}|}\begin{vmatrix} \mathbb{W} & \mathbb{Y}_1^{\dag} \\
\mathbb{Y}_3 & 0 \end{vmatrix} \end{pmatrix} \end{aligned}
\end{equation}%

where
\begin{equation}
\begin{aligned} \label{multi-bb-p} &\mathbb{W}=\begin{pmatrix}
\mathbb{W}_{1,1} & \cdots & \mathbb{W}_{1,n} \\ \vdots & \ddots & \vdots \\
\mathbb{W}_{n,1} & \cdots & \mathbb{W}_{n,n} \end{pmatrix},\\
&(\mathbb{W}_{j,k})_{\jmath,\ell}=\sum_{\chi=0}^{\jmath+\ell-2}
\sum_{\eta=\max{(0,\chi-\jmath+1)}}^{\min{\ell-1,\chi}}
\left(\frac{1}{\lambda_k-\lambda_j}\right)^{\chi+1}\times\\ &
C_\chi^{\eta}(\lambda_j^*)^{\chi-\eta}(-\lambda_k)^\eta
\Phi_{j,\jmath-1-\chi+\eta}^{\dag}\Xi\Phi_{k,\ell-1-\eta}, k=1,2,\ldots,n,\\
&\Phi(\lambda_k(1+\epsilon_k))=\Phi_{k,0}+\Phi_{k,1}\epsilon+\ldots+%
\Phi_{k,m_k}\epsilon^{m_k}+\ldots,\\ &\mathbb{Y}=(\mathbb{H}_1
,\mathbb{H}_2,\ldots,\mathbb{H}_n),
\mathbb{H}_k=(\Phi_{k,0},\Phi_{k,1},\ldots,\Phi_{k,m_k}). \end{aligned}
\end{equation}%
The term $(\mathbb{W}_{j,k})_{\jmath ,\ell }$ denotes the element located in
the $\jmath ^{th}$ row and $\ell ^{th}$ column of matrix $\mathbb{W}_{j,k}$,
which is of size $(m_{j}+1)\times (m_{k}+1)$, $\mathbb{Y}_{j}$ represents
the $j^{th}$ row of $\mathbb{Y}$, $\Phi (\lambda )$ are eigenfunction
solutions of the Lax pair (\ref{A1}) for the Manakov system~(\ref{Manakov}),
and $\Phi _{k,m_{k}}$ is the coefficient of the Taylor expansion of $\Phi
(\lambda _{k}(1+\epsilon ^{k}))$ at $\epsilon ^{k}=0$. Here, $n$ denotes the
order of the MP soliton/breather, while $m_{k}$ is the pole multiplicity of $%
m_{k}+1$ for these modes and $\lambda_j$ is the $j^{th}$ complex spectral parameter.

\section{Multi-pole solitons with the striped phase on zero background}

The zero-seed initialization, with $u_{10}=u_{20}=0$, enables the generation
of vector bright-bright MP stripe solitons with zero background (therefore
these solutions are categorized as bright ones). The Lax pair~(\ref{A1})
produces the respective elementary eigenfunction solution
\begin{equation}
\begin{aligned} \label{eigenfunction}
\boldsymbol{\Phi}(\lambda)=(e^{\theta},\beta_1e^{-\theta},\beta_2e^{-%
\theta})^T, \end{aligned}
\end{equation}%
with $\theta =i\lambda (x+\lambda t)$, $\beta _{1}$ and $\beta _{2}$ being
two real constants, and $\lambda $ representing the complex spectral
parameter.

\subsection{Double-pole (DP) solutions}

For the case of $n=1$ and $m_{k}=1$ in general solutions~(\ref{multi-pole}),
the bright-bright DP stipe-soliton solution with two pseudospin components $%
\Psi _{j}$ $(j=1,2)$ is%
\begin{equation}
\begin{aligned} \label{double-bb}
\Psi_j=&8\lambda_Ie^{-2i\theta_I}P_jB_2(x,t), \end{aligned}
\end{equation}%
where the helicoidal SOC-induced striped modulations are represented by $x$%
-periodic functions
\begin{equation}
\begin{aligned} \label{periodic-Pj} &P_1=e^{-i\kappa x}(\beta_1\nu_+e^{-ik_m
x}+\beta_2\nu_-e^{ik_m x}),\\ &P_2=e^{i\kappa x}(\beta_1\nu_-e^{-ik_m
x}-\beta_2\nu_+e^{ik_m x}). \end{aligned}
\end{equation}%
and the evolution of the DP solitons is governed by the semi-rational
function
\begin{equation}
\begin{aligned} \label{double-bb-B2}
B_2=\frac{\beta(-i+4\lambda_I^2t-i\theta_R)e^{-\theta_R}+
(-i+4\lambda_I^2t+i\theta_R)e^{\theta_R}}
{e^{2\theta_R}+\beta^2e^{-2\theta_R}+2\beta(1+32\lambda_I^4t^2+2%
\theta_R^2)}, \end{aligned}
\end{equation}%
with
\begin{equation}
\theta _{R}=-2\lambda _{I}(x+2\lambda _{R}t),~\theta _{I}=\lambda
_{R}x+(\lambda _{R}^{2}-\lambda _{I}^{2})t,~\beta =s(\beta _{1}^{2}+\beta
_{2}^{2}).  \label{parameters}
\end{equation}
Hereafter, subscripts $R$ and $I$ refer to real and imaginary parts of
complex parameters.

We now aim to rigorously investigate the propagation dynamics of the bright
DP stripe solitons encoded in solution~(\ref{double-bb}). On top of the
vanishing background, this bright soliton exhibits \textit{locally} periodic
stripe structures along the spatial $x$-direction, described by
\begin{equation}
\begin{aligned} \label{stripe-bb}
&P_{11}=|P_1|^2=\beta_1^2\nu_+^2+\beta_2^2\nu_-^2
+2\beta_1\beta_2\nu_+\nu_-\cos{(2k_mx)},\\
&P_{21}=|P_2|^2=\beta_1^2\nu_-^2+\beta_2^2\nu_+^2
-2\beta_1\beta_2\nu_+\nu_-\cos{(2k_mx)}, \end{aligned}
\end{equation}%
(see Fig. \ref{fig1} below), which differs from the spatiotemporally
periodic stripe structures on non-zero backgrounds which are addressed
below. The linear spectrum of system~(\ref{helicoidal
SOC}) exhibits identical minima at $\pm k_{\text{m}}$~\cite{Kartashov2017}, and their linear superposition results in the formation of periodic stripe structures. An asymptotic analysis reveals nonstationary nature of these DP solitons,
which, unlike conventional solitons, propagate along curved space-time
trajectories. This is explained by the balance between $\ t$ and terms $%
e^{\pm \theta _{R}}$ in Eq.~(\ref{double-bb-B2}). Our asymptotic analysis
therefore focuses on the $t\varpropto e^{\pm \theta _{R}}$ scaling relation:
\begin{subequations}
\begin{align}
& S_{j}^{1+}=\frac{2\lambda _{I}e^{-2i\theta _{I}}}{\sqrt{\beta }}P_{j}\text{%
sech}{[\theta _{R}-\ln {(8\sqrt{\beta }\lambda _{I}^{2}t)}]},
\label{asymptotic-double} \\
& ~~~~~~~~~~~~~~~~~(t\varpropto e^{\theta _{R}},t\rightarrow +\infty ,\theta
_{R}\rightarrow +\infty ),  \notag \\
& S_{j}^{1-}=\frac{-2\lambda _{I}e^{-2i\theta _{I}}}{\sqrt{\beta }}P_{j}%
\text{sech}{\left[ \theta _{R}+\ln {(-\frac{8\lambda _{I}^{2}t}{\sqrt{\beta }%
})}\right] }, \\
& ~~~~~~~~~~~~~~~~~(t\varpropto e^{-\theta _{R}},t\rightarrow -\infty
,\theta _{R}\rightarrow -\infty ),  \notag \\
& S_{j}^{2+}=\frac{2\lambda _{I}e^{-2i\theta _{I}}}{\sqrt{\beta }}P_{j}\text{%
sech}{\left[ \theta _{R}+\ln {(\frac{8\lambda _{I}^{2}t}{\sqrt{\beta }})}%
\right] }, \\
& ~~~~~~~~~~~~~~~~~(t\varpropto e^{-\theta _{R}},t\rightarrow +\infty
,\theta _{R}\rightarrow -\infty ),  \notag \\
& S_{j}^{2-}=\frac{-2\lambda _{I}e^{-2i\theta _{I}}}{\sqrt{\beta }}P_{j}%
\text{sech}{[\theta _{R}-\ln {(-8\sqrt{\beta }\lambda _{I}^{2}t)}]}, \\
& ~~~~~~~~~~~~~~~~~(t\varpropto e^{\theta _{R}},t\rightarrow -\infty ,\theta
_{R}\rightarrow +\infty ).  \notag
\end{align}%
Here, $S_{j=1,2}^{n\pm }$ represents asymptotic solitons in the $j$-th component,
superscript $n=1,2$ refers to different soliton branches, and $\pm$ indicates
the asymptotic state after and prior to the interaction, respectively.

The above asymptotic analysis confirms the amplitude conservation
\end{subequations}
\begin{equation}
\begin{aligned} \label{amplitude-bb} &|A_1^j|^2=4\lambda_I^2P_{11}/\beta,\\
&|A_2^j|^2=4\lambda_I^2P_{21}/\beta \end{aligned}
\end{equation}%
across the soliton collisions, where $A_{1}^{j}$ and $A_{2}^{j}$ represent the
amplitude of the $j$-th soliton in the first and second components,
respectively. With phase shifts between DP stripe solitons $\Psi _{j}^{1+}$
and $\Psi _{j}^{1-}$ (or between $\Psi _{j}^{2+}$ and $\Psi _{j}^{2-}$)
being $\delta _{1,2}=\mp 2\ln {(8\lambda _{I}^{2}|t|)}$, respectively, the
DP stripe soliton collisions are fully elastic, which is quite natural for
the integrable system.

In the asymptotic regime, the soliton's central trajectories, along which
the amplitudes attain their maxima, take the form of
\begin{equation}
\begin{aligned} \label{trajectory-bb}
&S_j^{1+}:~e^{\theta_R}-8\sqrt{\beta}\lambda_I^2t=0;~
S_j^{1-}:~e^{-\theta_R}+\frac{8\lambda_I^2t}{\sqrt{\beta}}=0;\\
&S_j^{2+}:~e^{-\theta_R}-\frac{8\lambda_I^2t}{\sqrt{\beta}}=0;~
S_j^{2-}:~e^{\theta_R}+8\sqrt{\beta}\lambda_I^2t=0. \end{aligned}
\end{equation}
In addition, the trajectories' temporal slopes are given by
\begin{equation}
\begin{aligned} \label{slope-bb}
&S_j^{1+}:-2\lambda_R-\frac{1}{2\lambda_It}(t>0);~
S_j^{1-}:-2\lambda_R+\frac{1}{2\lambda_It}(t<0);\\
&S_j^{2+}:-2\lambda_R+\frac{1}{2\lambda_It}(t>0);~
S_j^{2-}:-2\lambda_R-\frac{1}{2\lambda_It}(t<0). \end{aligned}
\end{equation}%
The magnitude of the asymptotic soliton's slope can be used to determine its
distribution around the line $\mathcal{L}$, defined as $\theta
_{R}=-2\lambda _{I}(x+2\lambda _{R}t)=0$. Moreover, it can be observed that,
at $|t|\rightarrow \infty $, the shape of the asymptotic soliton gradually
becomes parallel to line $\mathcal{L}$, with a limiting slope of $-2\lambda
_{R}$.

In Fig.~\ref{fig1}, we display the density distribution in the static and
moving bright DP stripe solitons, alongside their asymptotic trajectories
governed by Eq.~(\ref{trajectory-bb}). The results corroborate the agreement
between the asymptotic predictions and full soliton evolutions in both
cases. These solitons are characterized not only by the curved trajectories
--- a hallmark of the DP dynamics --- but also exhibit periodic stripe
modulations along in the spatial direction under the action of the
helicoidal SOC, justifying their designation as DP stripe solitons. Note
that, while the dual pseudospin components share identical curved
trajectories, they manifest distinct amplitude profiles and out-of-phase
configurations governed by Eqs.~(\ref{amplitude-bb}) and (\ref{stripe-bb}),
respectively, which is a direct consequence of the symmetry breaking induced
by the helicoidal SOC.

\begin{figure}[th]
\includegraphics[scale=0.4]{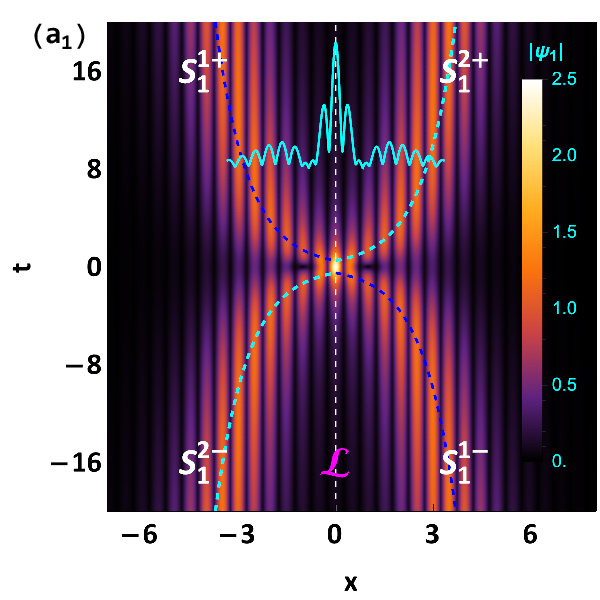}\hspace{3mm} %
\includegraphics[scale=0.4]{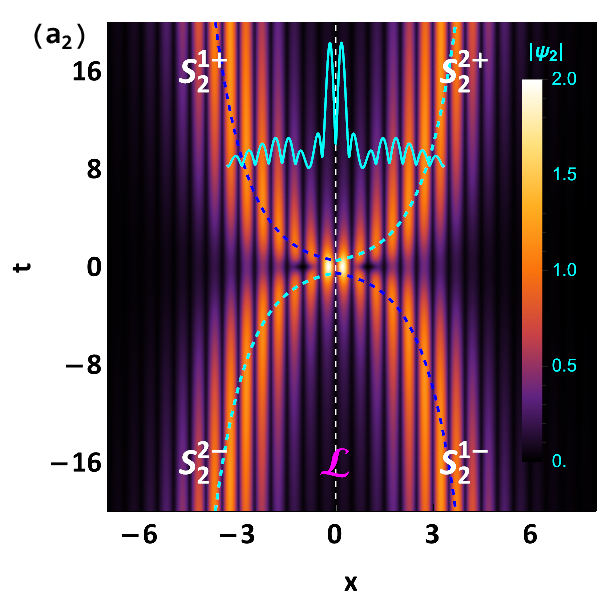}\newline
\includegraphics[scale=0.4]{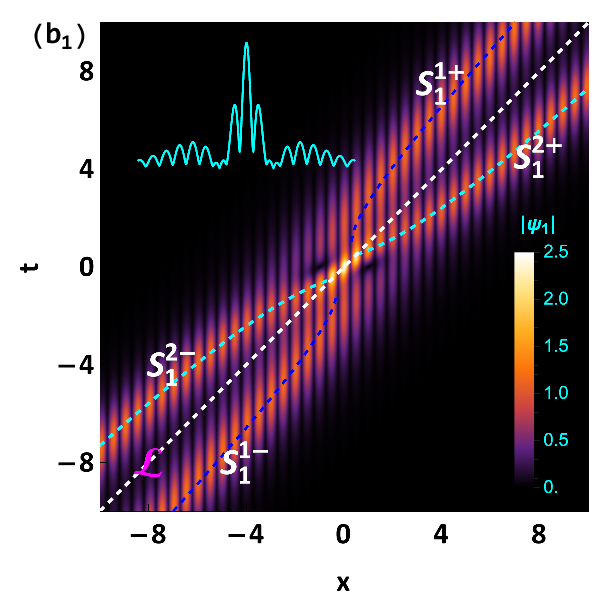}\hspace{3mm} %
\includegraphics[scale=0.4]{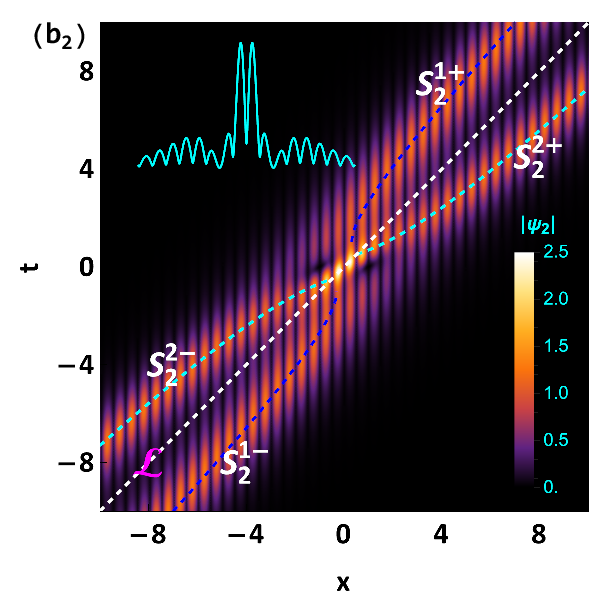}\newline
\vspace{-3mm}
\caption{($a$) Static and ($b$) moving bright DP stripe solitons, where
dashed curves map asymptotic soliton paths and line $\mathcal{L}$, defined
as $x+2\protect\lambda _{R}t=0$, determines their convergence direction. The
parameter sets are: ($a$) $\protect\lambda _{R}=0$ and ($b$) $\protect%
\lambda _{R}=-0.5$. Other parameters are $s=1$, $\protect\alpha =6$, $%
\protect\kappa =2$, $\protect\beta _{1}=\protect\beta _{2}=\frac{\protect%
\sqrt{2}}{2}$ and $\protect\lambda _{I}=0.6$. The cyan curves illustrate the
out-of-phase configurations between the two-component waveforms at $t=0$.}
\label{fig1}
\end{figure}

\begin{figure}[th]
\includegraphics[scale=0.7]{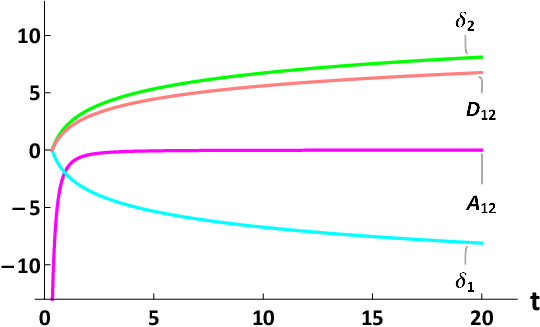}\newline
\vspace{-3mm}
\caption{The time evolution of phase shifts $\protect\delta _{1,2}$,
relative distance $D_{12}$, and separation acceleration $A_{12}$ between the
DP asymptotic solitons. The parameters are the same as in Fig.~\protect\ref%
{fig1}(a).}
\label{fig1a}
\end{figure}

Another characteristic feature of the DP solitons is the logarithmic
temporal variation of the separation between the two solitons. The
asymptotic analysis demonstrates that the distance between asymptotic
solitons $\Psi _{j}^{(1)}$ and $\Psi _{j}^{(2)}$ is $D_{12}=|\lambda
_{I}|^{-1}\ln {(8\lambda _{I}^{2}|t|)}$, growing logarithmically with time.
We stress that the separation acceleration $A_{12}=-64|\lambda _{I}|^{3}\exp
\left( -2|\lambda _{I}|D_{12}\right) $, which is calculated as the second
time derivative of $D_{12}$, exhibits exponential decay as the function of
the separation, being a stark departure from conventional behavior for
solitons, \textit{viz}., $D\propto t$. The evolution of the phase shifts $%
\delta _{1,2}$, relative distance $D_{12}$, and separation acceleration $%
A_{12}$ between the asymptotic solitons are shown in Fig.~\ref{fig1a}. It
demonstrates that the interaction force is strong during the collision,
resulting in the significant acceleration. At the post-collision stage, as
the separation between the solitons increases, the interaction force
gradually decays. The interaction force and acceleration asymptotically
vanish at $t\rightarrow \infty $, as the solitons become infinitely
separated.

The formation of the stripe state of the DP solitons is determined by the
helicoidal SOC. The period along the $x$ direction is $T_{x}=\pi /\sqrt{%
\alpha ^{2}+\kappa ^{2}}$, which indicates that the stripe period is
determined by the strength and frequency of the helicoidal SOC. The
consideration of the asymptotic amplitudes (see Eq.~(\ref{amplitude-bb}))
reveals that the solitons in the same pseudospin component exhibit identical
amplitudes and stripe periods, while inter-component density distributions $%
P_{11}$ and $P_{21}$ display explicit dependence on $\alpha $ and $\kappa $.
Notably, the total density $|\mathbf{\Psi }|^{2}=\mathbf{\Psi }^{\dag }%
\mathbf{\Psi }$ is nonperiodic along $x$, contrasting with the
intra-component periodic structures.

The amplitude/intensity of the DP solitons in the interaction region can be
controlled by the helicoidal SOC. At the origin point $(0,0)$, the
cross-component coupling generates amplified wave peaks described by
\begin{equation}
\begin{aligned} \label{amplitude00-bb}
(|\Psi_1|^2,|\Psi_2|^2)|_{(0,0)}=8\lambda_I^2[f_1(\alpha,\kappa),f_2(\alpha,%
\kappa)], \end{aligned}
\end{equation}%
where the amplitude adjustment factors $f_{1}$ and $f_{2}$ are
\begin{equation}
\begin{aligned} \label{amplitude00-bb-2}
f_1=1+\frac{\alpha}{\sqrt{\alpha^2+\kappa^2}},~
f_2=1-\frac{\alpha}{\sqrt{\alpha^2+\kappa^2}}. \end{aligned}
\end{equation}%
Note that at $\alpha =0$ (the Manakov-system limit), the peak amplitudes of
the colliding wave components at their respective collision centers exhibit
both equality and constancy of these factors, $f_{1}=f_{2}=1$. These peak
amplitudes demonstrate significant sensitivity to parameters of the
helicoidal SOC, sharply diverging from the amplitude uniformity observed in
the SOC-free systems. This parametric tunability, combined with the
geometric constraints imposed by $k_{m}$, highlights the unique interplay
between the SOC physics and nonlinear wave dynamics in the two-component
soliton systems.

The mirror-symmetric chart of the amplitude modulation factors in Fig.~\ref%
{fig2} reflects the energy redistribution mechanism in the $(\alpha ,\kappa )
$ parameter space. This symmetry enforces an anti-correlated relationship:
the soliton amplification in either component, represented by the factors $%
f_{1}$ and $f_{2}$, is accompanied by the suppression in its counterpart,
preserving the total amplitude invariant, $f_{1}+f_{2}=2$. The
reflection symmetry with respect to $\kappa \leftrightarrow -\kappa$ implies
that the amplitude amplification or attenuation is independent of the (right- or left-handed)
helicity, whereas the finite SOC strength ($\alpha \neq 0$)
breaks the Manakov-system's degeneracy (the dashed line corresponding to $%
\alpha =0$), enforcing the component-selective amplification or attenuation.

\begin{figure}[th]
\includegraphics[scale=0.35]{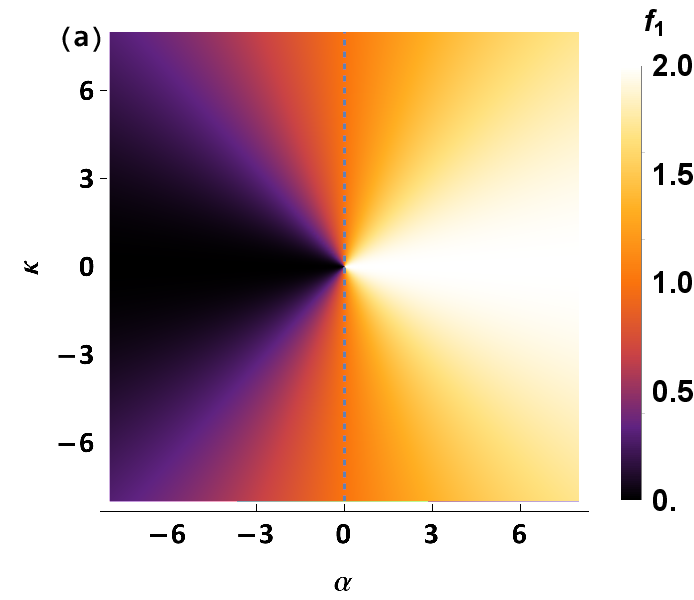}\hspace{3mm} %
\includegraphics[scale=0.35]{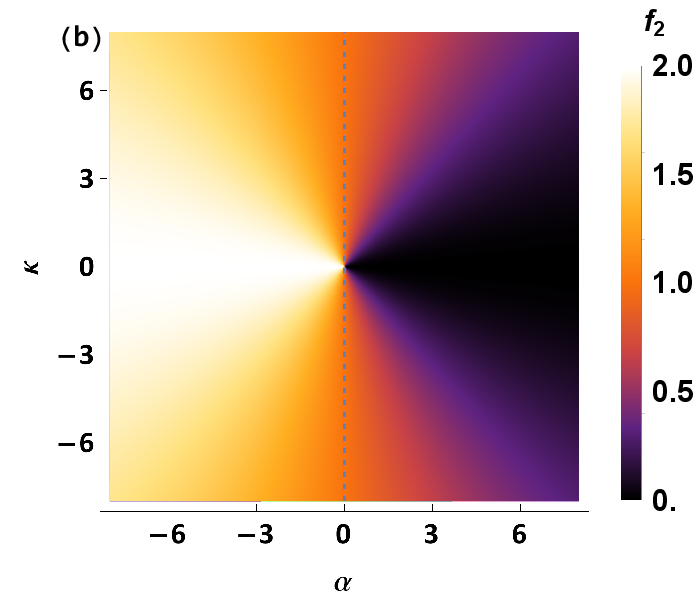}\newline
\vspace{-3mm}
\caption{The mirror-symmetric distribution of the amplitude modulation
factors $f_{1}$ and $f_{2}$ in the $(\protect\alpha ,\protect\kappa )$%
-plane, with the dashed lines denoting the degenerate Manakov case $(\protect%
\alpha =0)$, with $f_{1,2}\equiv1$.}
\label{fig2}
\end{figure}

\subsection{Triple-pole solutions}

Solution~(\ref{multi-pole}) with $n=1$ and $m_{k}=2$ under the zero-seed
initialization produces bright-bright solitons with the triple-pole
characteristics and stripe modulation, as given by
\begin{equation}
%\begin{aligned}
\label{triple-bb}
\Psi_j=%&
-4i\lambda_Ie^{-2i\theta_I}P_jB_3(x,t).
%\end{aligned}
\end{equation}%
The stripe-modulating functions $P_{j}$ given by Eq.~(\ref{stripe-bb})
remain the same for solutions with higher pole numbers, while the
semi-rational function $B_{3}$ describing the triple-pole configuration is
\begin{widetext}
\begin{equation}
\label{triple-bb-b3}
B_3=\frac{(-32\lambda_I^4t^2+\Delta_1)e^{2\theta_R}+\beta^2
(-32\lambda_I^4t^2+\Delta_2)e^{-2\theta_R}-2\beta\Delta_3}
{e^{3\theta_R}+\beta^3e^{-3\theta_R}+\beta^2(1024\lambda_I^8t^2+\Gamma_2)e^{-\theta_R}
+\beta(1024\lambda_I^8t^4+\Gamma_1)e^{\theta_R}},
\end{equation}
\end{widetext}where $\theta _{R}$, $\theta _{I}$ and $\beta $ preserve their
definitions given by Eq.~(\ref{parameters}), with the other parameters
defined in Appendix B. This hierarchical construction can extend the DP
formalism to higher-order pole solutions, while maintaining the same stripe
modulation $P_{j}$.

The asymptotic behavior of the triple-pole solitons, governed by the
semi-rational structure of $B_{3}$, reveals a geometric dichotomy between
asymptotically straight and curved trajectories, which was absent in the DP
case. Two linear asymptotic solitons are confined to the critical line $\mathcal{L}$, $%
\theta _{R}-\ln \sqrt{\beta }=-2\lambda _{I}(x+2\lambda _{R}t)-\ln \sqrt{%
\beta }=0$, exhibiting universal profiles
\begin{equation}
\begin{aligned} \label{asymptotic-triple-1}
&S_j^{2\pm}=\frac{2i\lambda_Ie^{-2i\theta_I}}{\sqrt{\beta}}P_j
\text{sech}{(\theta_R-\ln{\sqrt{\beta}})},\\ &~~~~~~(t\rightarrow \pm\infty,
\theta_R-\ln{\sqrt{\beta}}=\textit{const}), \end{aligned}
\end{equation}%
while four curved asymptotic solitons emerge from the balances between $t$
and $e^{\pm \theta _{R}/2}$, with the trajectories governed by
\begin{equation}
\begin{aligned} \label{asymptotic-triple-2}
&S_j^{1+}=S_j^{3-}=\frac{2i\lambda_Ie^{-2i\theta_I}}{\sqrt{\beta}}P_j
\text{sech}{[\theta_R-\ln{(32\sqrt{\beta}\lambda_I^4t^2)}]},\\
&~~~~~~~~~~~~~~~~~(t\varpropto e^{\theta_R/2}, \theta_R\rightarrow +\infty,
t\rightarrow \pm\infty),\\
&S_j^{3+}=S_j^{1-}=\frac{2i\lambda_Ie^{-2i\theta_I}}{\sqrt{\beta}}P_j
\text{sech}{(\theta_R+\ln{\frac{32\lambda_I^4t^2}{\sqrt{\beta}}})},\\
&~~~~~~~~~~~~~~~~~(t\varpropto e^{-\theta_R/2}, \theta_R\rightarrow -\infty,
t\rightarrow \pm\infty). \end{aligned}
\end{equation}%
The separation of the asymptotic trajectories into the linear and curved
types highlights the intrinsic relationship between\ the pole multiplicity
and soliton kinematics, where the asymptotic expressions~(\ref%
{asymptotic-triple-1}) and (\ref{asymptotic-triple-2}) represent the energy
redistribution between geometrically distinct soliton branches. Note that
the constraint admitting the existence of the straight asymptotic
trajectories $\mathcal{L}$ reflects the system's residual symmetry, whereas
the curved trajectories manifest broken translational invariance through
their $\theta _{R}$-dependent path deformations.

The asymptotic analysis conclusively demonstrates that the interactions
among triple-pole stripe solitons remain strictly elastic. This conclusion
is upheld by the invariant amplitude relationships
\begin{equation}
\begin{aligned} \label{amplitude-bb-3}
&|A_1^j|^2=4\lambda_I^2P_{11}/\beta,\\ &|A_2^j|^2=4\lambda_I^2P_{21}/\beta,
(j=1,2,3) \end{aligned}
\end{equation}%
accompanied by certain phase shifts. The first ($S_{j}^{1}$) and third ($%
S_{j}^{3}$) asymptotic solitons acquire phase shifts $\delta _{1,3}=\mp 4\ln
{(4\sqrt{2}\lambda _{I}^{2}|t|)}$ respectively, while the second ($S_{j}^{2}$%
) asymptotic soliton exhibits zero phase shift $\delta _{2}=0$, due to its
configuration which is collinear to the interaction axis.

The central trajectories of the asymptotic solitons for the triple-pole
stripe solitons can be obtained from the above asymptotic analysis as
\begin{equation}
\begin{aligned} \label{trajectory-bb-3} &S_j^{1+},
S_j^{3-}:~e^{\theta_R}-32\sqrt{\beta}\lambda_I^4t^2=0;\\ &S_j^{1-},
S_j^{3+}:~e^{-\theta_R}+\frac{32\lambda_I^4t^2}{\sqrt{\beta}}=0;\\
&S_j^{2\pm}: \theta_R-\ln{\sqrt{\beta}}=0, \end{aligned}
\end{equation}%
at which the soliton's amplitude attains its maximum. Furthermore, the
time-dependent slopes of the trajectories can be derived as
\begin{equation}
\begin{aligned} \label{slope-bb-3} &S_j^{2+}:-2\lambda_R(t>0);~
S_j^{2-}:-2\lambda_R(t<0);\\
&S_j^{1+}:-2\lambda_R-\frac{1}{\lambda_It}(t>0);~
S_j^{1-}:-2\lambda_R+\frac{1}{\lambda_It}(t<0);\\
&S_j^{3+}:-2\lambda_R+\frac{1}{\lambda_It}(t>0);~
S_j^{3-}:-2\lambda_R-\frac{1}{\lambda_It}(t<0). \end{aligned}
\end{equation}%
It can be observed that, at $t\rightarrow \pm \infty $, the curved
asymptotic solitons $S_{j}^{1,3\pm }$ asymptotically align parallel to the
linear ones $S_{j}^{2\pm }$, with a limiting slope of $-2\lambda _{R}$.

\begin{figure}[th]
\includegraphics[scale=0.34]{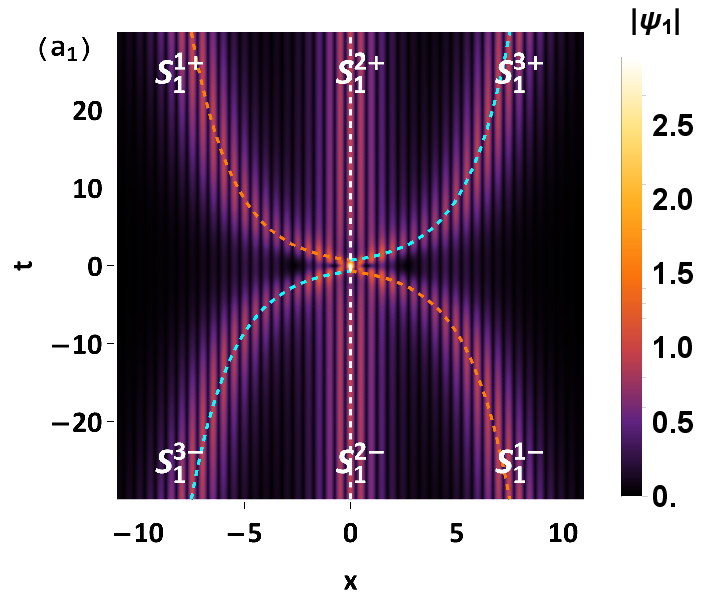}\hspace{3mm} %
\includegraphics[scale=0.34]{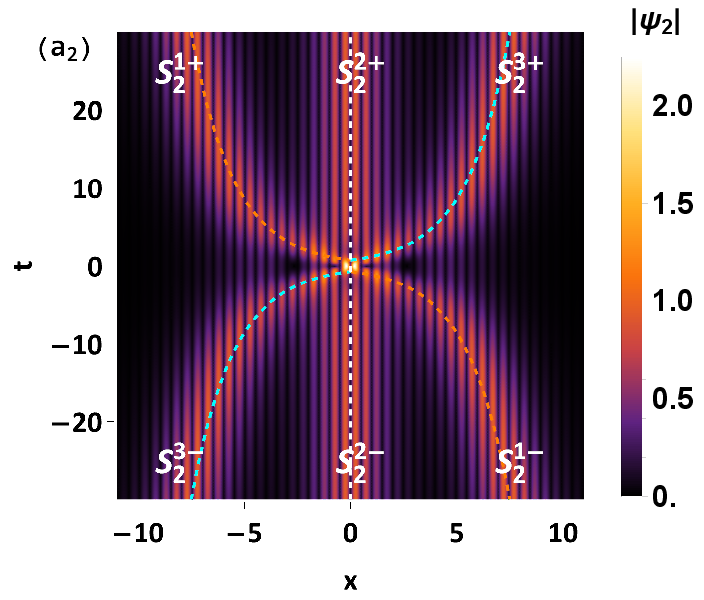}\newline
\includegraphics[scale=0.35]{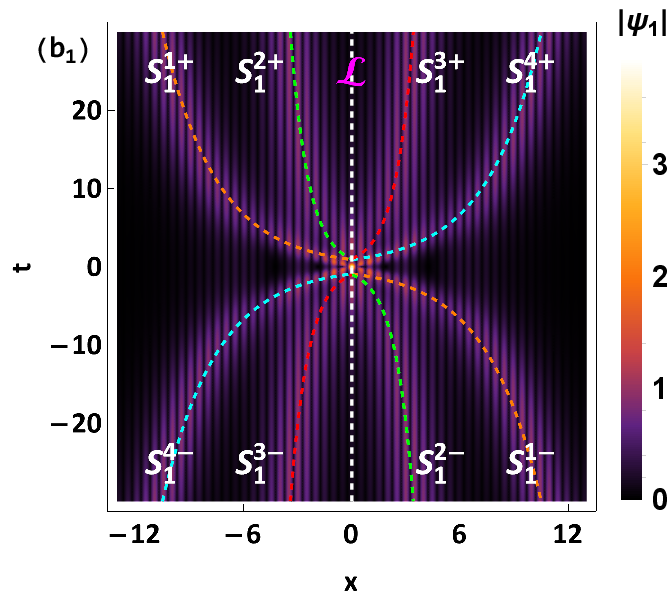}\hspace{3mm} %
\includegraphics[scale=0.35]{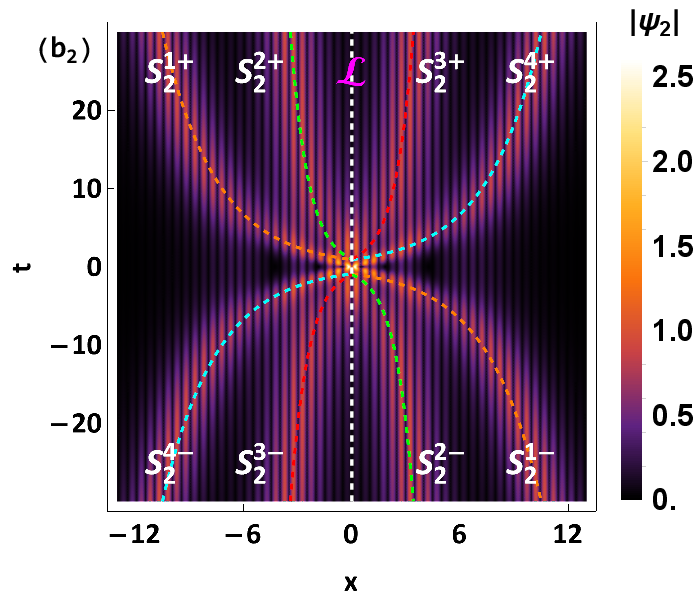}\newline
\includegraphics[scale=0.55]{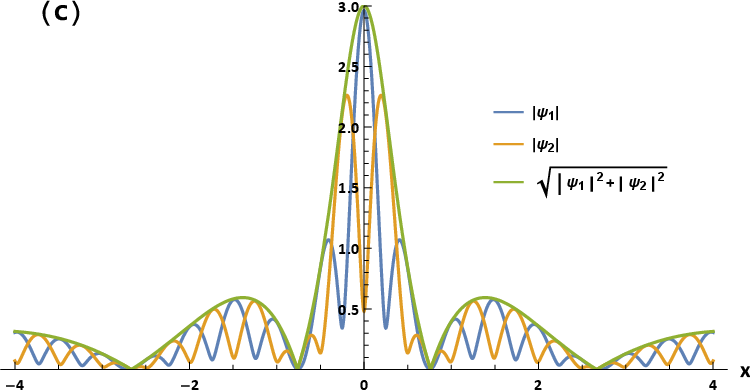}\newline
\vspace{-3mm}
\caption{Static bright triple-pole (a) and quadruple-pole (b) stripe
solitons, with the dashed lines showing the asymptotic soliton trajectories.
(c) The out-of-phase stripe structure of the triple-pole solitons at $t=0$.
Parameter are $s=1$, $\protect\lambda _{I}=0.5$, $\protect\lambda _{R}=0$, $%
\protect\alpha =6$, $\protect\kappa =2$, and $\protect\beta _{1}=\protect%
\beta _{2}=1/\protect\sqrt{2}$.}
\label{fig3}
\end{figure}

Figure~\ref{fig3}(a) presents the density distribution and asymptotic
trajectories of stationary triple-pole stripe solitons, where the
analytically predicted trajectories from Eq.~(\ref{trajectory-bb-3})
demonstrate remarkable consistency with the numerically found density
distribution. Compared to their DP counterparts, the triple-pole solitons
exhibit two distinct features: (i) curved asymptotic solitons with modified
curvature parameters, and (ii) two additional parallel linear asymptotic
solitons absent in DP cases, while maintaining similar stripe structures.
Simultaneously presented in Fig.~\ref{fig3}(c) is the out-of-phase stripe
structure of the triple-pole solitons in the two-component system, captured
at the initial time $t=0$.

\begin{figure}[th]
\includegraphics[scale=0.8]{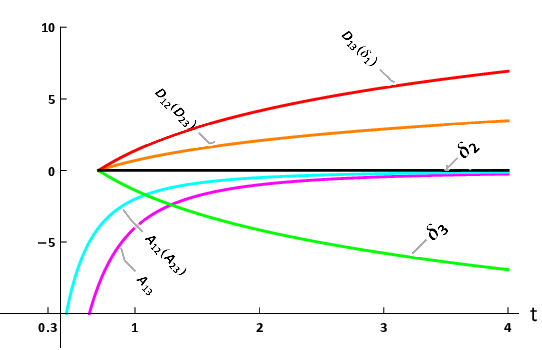}\newline
\vspace{-3mm}
\caption{The evolution of the phase shifts $\protect\delta _{1,2,3}$,
relative distance $D_{ij}$ ($i,j=1,2,3$), and separation acceleration $%
A_{ij} $ between the triple-pole asymptotic solitons. The parameters are the
same as in Fig.~\protect\ref{fig3}(a).}
\label{fig3a}
\end{figure}

For the triple-pole stripe solitons, the inter-soliton spacings $%
D_{12}=D_{23}=\frac{1}{2}D_{13}=|\lambda _{I}|^{-1}\ln {(4\sqrt{2}\lambda
_{I}^{2}|t|)}$ for solitons $S_{j}^{1,2,3}$ in each component follow the
logarithmic time dependence. The corresponding acceleration dynamics reveal
exponentially decaying profiles $A_{12}=A_{23}=-32|\lambda
_{I}|^{3}e^{-2|\lambda _{I}|D_{12}}$ and $A_{13}=-64|\lambda
_{I}|^{3}e^{-|\lambda _{I}|D_{13}}$, with the decay coefficient $\lambda
_{I} $. The exponential suppression of the acceleration with the increase of
the separation distance reproduces the behavior which was exhibited above by
the DP solitons. The evolution of the phase shifts $\delta _{1,2,3}$,
relative distances $D_{ij}$ ($i,j=1,2,3$), and separation acceleration $%
A_{ij}$ between the triple-pole asymptotic solitons are displayed in Fig.~%
\ref{fig3a}. It seen that, similar to DP solitons, the triple-pole ones
exhibit strong interaction forces and significant acceleration of the
separation between the curved solitons during collisions. As the propagation
proceeds, the logarithmic growth of the separation between the curved
solitons is the dominant feature, with the respective asymptotically
decaying acceleration.

Extending the analysis to quadruple-pole stripe solitons through Eq.~(\ref%
{multi-pole}) with parameters $n=1$ and $m_{k}=3$, Fig.~\ref{fig3}($b$)
illustrates their evolution, featuring eight asymptotic trajectories, while
cumbersome analytical expressions are omitted here. In fact, setting $n=1$
and $m_{k}=N-1$ ($N\geq 1$), the $N$-pole stripe solitons, produced by Eq.~(%
\ref{multi-pole}), exhibit universal structural characteristics: solitons
with odd-order pole numbers universally contain both curved asymptotic
components (with the curvature depending on the pole number) and central
straight solitons, a feature which is absent in even-order cases.

The above analysis facilitates deeper understanding of the dynamics and
interactions of the solitons, emphasizing their distinctive curved
trajectories which set them apart from the traditional multi-solitons and
bound-state solitons. The term "multi-pole" reflects the complexity and
hierarchical nature of these soliton states, highlighting their nonlinear
characteristics and dynamical behavior.

\section{Multi-pole solitons with beating stripe modes on the plane-wave
background}

Here we extend the consideration of MP solitons driven by the helicoidal SOC
to the case of nonzero background. The recently reported beating stripe
soliton formation, arising from the dark-bright soliton superposition \cite%
{ding2025}, is the motivation for constructing MP beating stripe solitons
through a superposition of MP dark and bright solitons. With the plane
wave-zero seed solutions $u_{10}=e^{i\theta _{1}}$ and $u_{20}=0$, Lax pair~(%
\ref{A1}) yields the elementary eigenfunction solution
\begin{equation}
\begin{aligned} \label{eigenfunction2}
\boldsymbol{\Phi}(\lambda)=\left(l_1e^{-iA(\mu_1)},
\frac{l_1e^{-i(A(\mu_1)-\theta_1)}}{k_1-\mu_1},l_3e^{-i(A(k_2)-\theta_2)}%
\right)^T, \end{aligned}
\end{equation}%
where
\begin{equation}
\begin{aligned} A(\xi)=&(\lambda+\xi)x+(s+\lambda^2-\frac{\xi^2}{2})t,\\
\mu_1=&-2\lambda_{R}-\sqrt{s-\lambda_{I}^2}-i\lambda_{I},\\
\theta_1=&k_jx+(s-\frac{1}{2}k_1^2)t, \end{aligned}
\end{equation}%
$k_{1}=-2\lambda _{R}$, and $l_{1}$ and $l_{3}$ are constant parameters.

Using eigenfunction~(\ref{eigenfunction2}) as the basis and applying
solutions~(\ref{multi-pole}) with $n=1$ and $m_{k}=1$, we obtain vector DP
beating stripe-soliton solutions in the form of
\begin{equation}
\begin{aligned} \label{Double-pole dark-bright} \Psi_1=&e^{-i\kappa
x}(\nu_+\Psi_{D}+\nu_-\Psi_{B}),\\ \Psi_2=&e^{i\kappa
x}(\nu_-\Psi_{D}-\nu_+\Psi_{B}), \end{aligned}
\end{equation}%
where the DP dark and bright solitons, $\Psi _{D}$ and $\Psi _{B}$, are
\begin{equation}
\begin{aligned}
\label{D-d-b}
\Psi_{D}=&e^{i\theta^\prime_1}-
\frac{8\lambda_{I}^3(2\lambda_{I}^2z^5st^2+\mathcal{D}_1)e^{i\theta^%
\prime_1}} {4\Lambda^2\lambda_{I}^6e^{2\delta}+\Lambda^{-2}\beta^2e^{-2\delta}
+8\lambda_{I}^3(\lambda_{1I}^2z^5st^2+\mathcal{D}_2)},\\
\Psi_{B}=&8\lambda_{I}e^{i\theta^\prime_2} \frac{2\Lambda
z\lambda_{I}^5(z^2t+i\delta-i)e^\delta
+\Lambda^{-1}\beta(z^3\lambda_{I}^2t-\mathcal{D}_3)e^\delta}
{4\Lambda^2\lambda_{I}^6e^{2\delta}+\Lambda^{-2}\beta^2e^{-2\delta}
+8\lambda_{I}^3(\lambda_{I}^2z^5st^2+\mathcal{D}_2)}, \end{aligned}
\end{equation}%
with
\begin{equation}
\begin{aligned} &\theta^\prime_1=\theta_1-k_mx,~~
\theta^\prime_2=\theta_1+z^2t/2+k_mx,\\
&\delta=-z(x+2\lambda_Rt),~~\beta=s\lambda_I(z\lambda_I-s),~~\Lambda=l_1/l_3.
\end{aligned}
\end{equation}%
Here, the Joukowsky transform,
\begin{equation}
\lambda _{I}=\left( z+s/z\right) /2,  \label{Jouk}
\end{equation}
is adopted to eliminate square-root complexities in $\mu _{1}$. For $s=1$,
the Joukowsky transformation confines $\lambda _{I}$ to the domain $|\lambda
_{I}|\geq 1$. To address the case of $|\lambda _{I}|\leq 1$, we implement
parameterization $\lambda _{I}=\sin \gamma $ $(-\pi /2\leq \gamma \leq \pi
/2)$, with explicit analytical forms of DP dark $\Psi _{D}$ and bright $\Psi
_{B}$ solitons derived under this mapping, as detailed in Appendix C.

Systematic asymptotic analysis reveals that these DP beating stripe solitons
inherit the trajectory curvature from the $t\varpropto e^{\pm \delta }$
scaling equilibrium, a property shared with the zero-background MP bright
solitons in Section III. The asymptotic solitons in the $|\lambda _{I}|\geq
1 $ case (the case of $|\lambda _{I}|<1$ is similar, therefore it is not
explicitly presented here) are explicitly constructed as:
\begin{subequations}
\begin{align}
& |S_{j}^{1+}|^{2}=\nu _{1j}+\nu _{2j}\text{sech}^{2}\Delta
_{1}-(-1)^{j}\Psi _{p}\tanh \Delta _{1}\text{sech}\Delta _{1},
\label{asymptotic-double-beating} \\
& ~~~~~~~~~~~~~~~~~(t\varpropto e^{\delta },t\rightarrow +\infty ,\delta
\rightarrow +\infty ),  \notag \\
& |S_{j}^{1-}|^{2}=\nu _{1j}+\nu _{2j}\text{sech}^{2}\Delta
_{2}-(-1)^{j}\Psi _{p}^{\prime }\tanh \Delta _{2}\text{sech}\Delta _{2}, \\
& ~~~~~~~~~~~~~~~~~(t\varpropto e^{-\delta },t\rightarrow -\infty ,\delta
\rightarrow -\infty ),  \notag \\
& |S_{j}^{2+}|^{2}=\nu _{1j}+\nu _{2j}\text{sech}^{2}\Delta
_{3}+(-1)^{j}\Psi _{p}^{\prime }\tanh \Delta _{3}\text{sech}\Delta _{3}, \\
& ~~~~~~~~~~~~~~~~~(t\varpropto e^{-\delta },t\rightarrow +\infty ,\delta
\rightarrow -\infty ),  \notag \\
& |S_{j}^{2-}|^{2}=\nu _{1j}+\nu _{2j}\text{sech}^{2}\Delta
_{4}+(-1)^{j}\Psi _{p}\tanh \Delta _{4}\text{sech}\Delta _{4}, \\
& ~~~~~~~~~~~~~~~~~(t\varpropto e^{\delta },t\rightarrow -\infty ,\delta
\rightarrow +\infty ),  \notag
\end{align}%
where
\end{subequations}
\begin{equation}
\begin{aligned} \label{asymptotic-p}
&\nu_{11}=\nu_+^2,~~\nu_{12}=\nu_-^2,~~\nu_{21}=-\nu_+^2+(S\nu_-)^2,\\
&\nu_{22}=-\nu_-^2+(S\nu_+)^2,~~\Psi_p^\prime=s~\text{Sgn}(z^2-s)\Psi_p,\\
&\Psi_p=2S\nu_+\nu_-\cos(2k_m x+z^2t/2),~~S=\sqrt{s(z^2+s)},\\
&\Delta_1=\delta-\log\left(\frac{2z^3t}{\Lambda S}\right),
~~\Delta_2=\delta+\log\left(\frac{-2\Lambda zt S^3} {|z^2-s|}\right),\\
&\Delta_3=\delta+\log\left(\frac{2\Lambda zt S^3} {|z^2-s|}\right),
~~\Delta_4=\delta-\log\left(\frac{-2z^3t}{\Lambda S}\right). \end{aligned}
\end{equation}

\begin{figure*}
\includegraphics[scale=0.45]{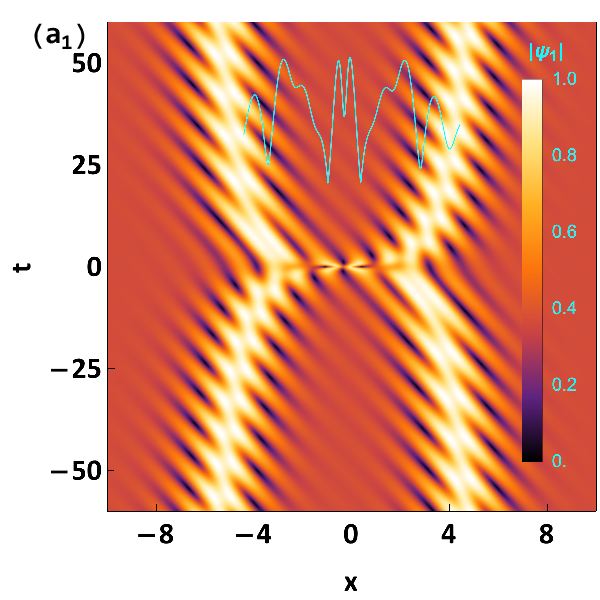}\hspace{3mm}
\includegraphics[scale=0.45]{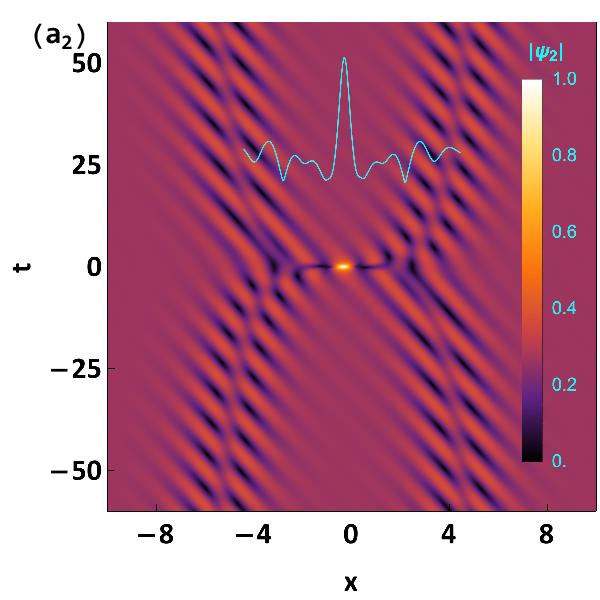} \hspace{3mm}
\includegraphics[scale=0.45]{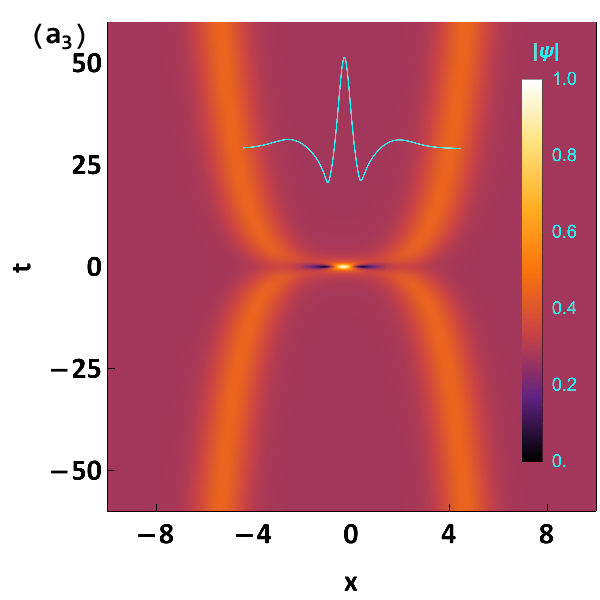} \hspace{3mm}\newline
\includegraphics[scale=0.45]{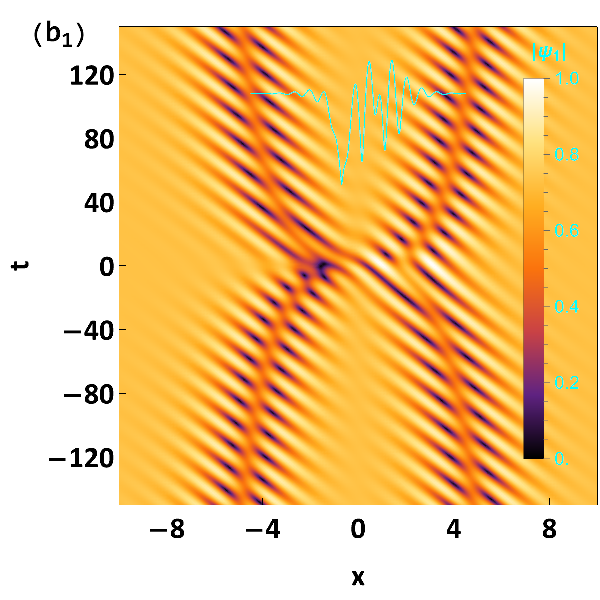}\hspace{3mm}
\includegraphics[scale=0.45]{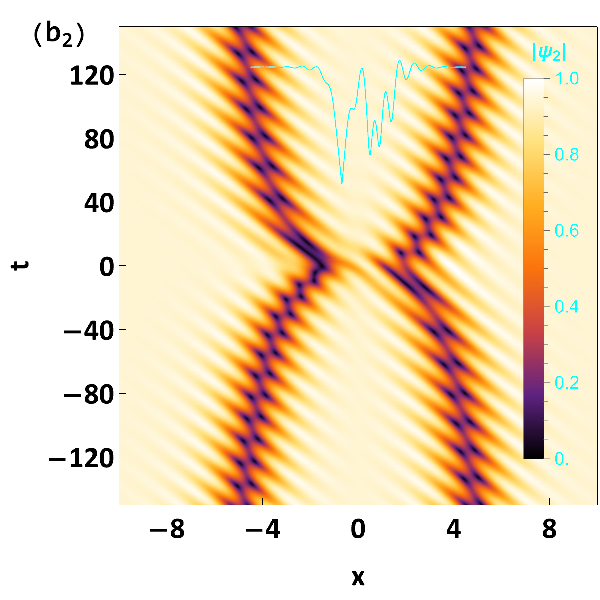} \hspace{3mm}
\includegraphics[scale=0.45]{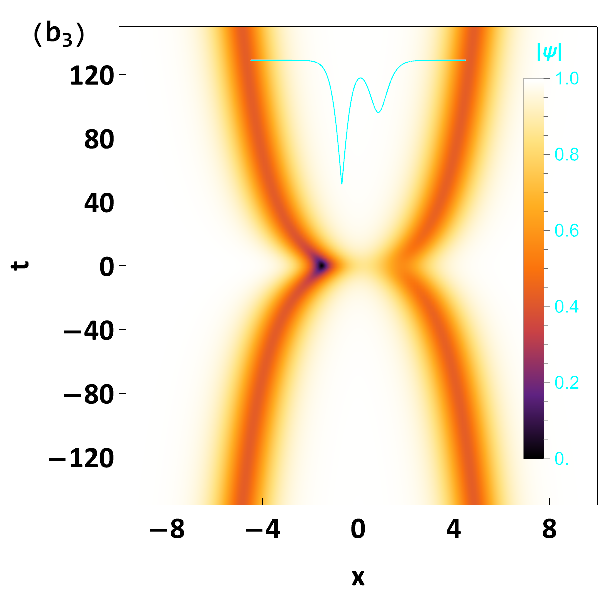} \hspace{3mm}\newline
\includegraphics[scale=0.45]{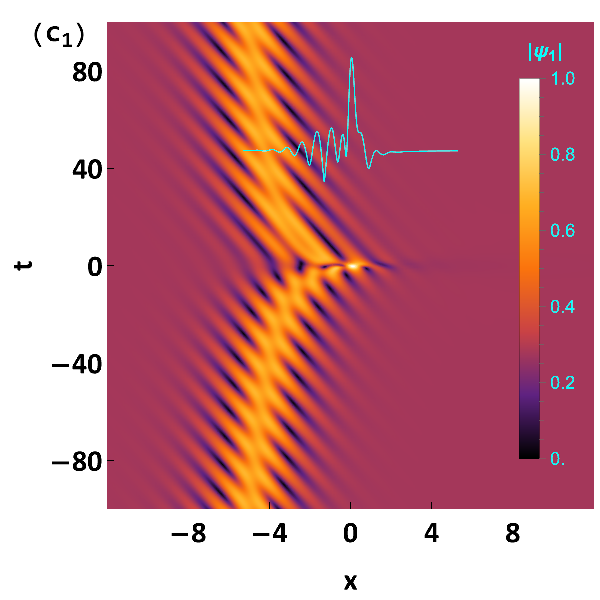}\hspace{3mm}
\includegraphics[scale=0.45]{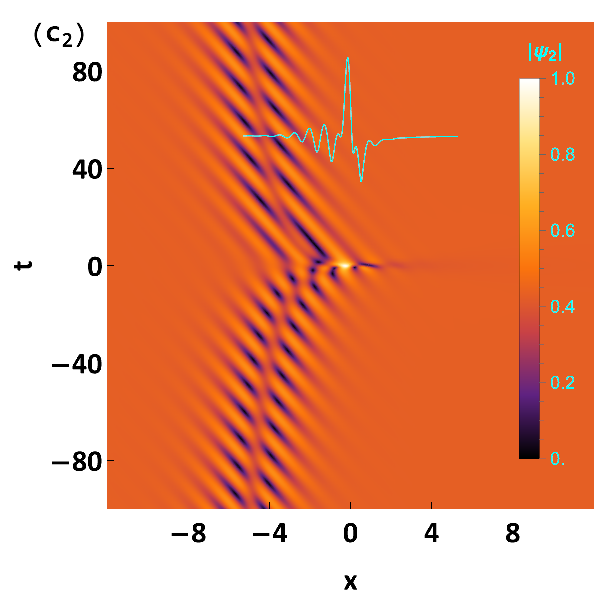} \hspace{3mm}
\includegraphics[scale=0.45]{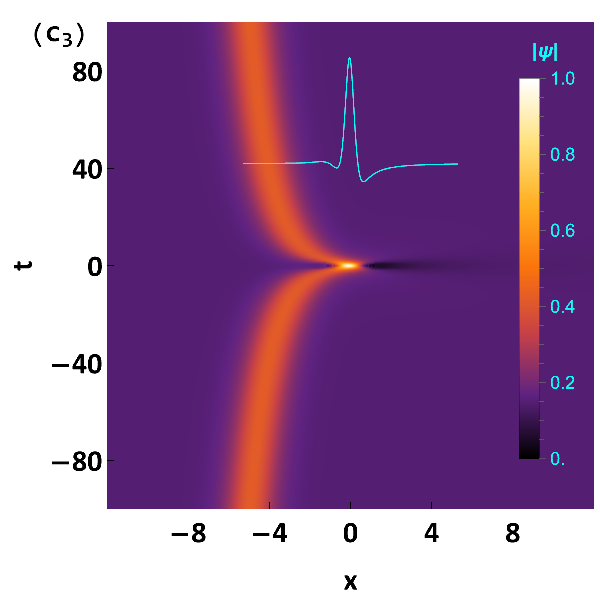} \hspace{3mm}\newline
\vspace{-3mm}
\caption{\label{fig5}Static DP beating stipe solitons in components $|\Psi _{1,2}|$ with
the bright (a) and dark (b) total density, $|\Psi |^{2}=|\Psi
_{1}|^{2}+|\Psi _{2}|^{2}$, in the case of the attractive and repulsive
interatomic interactions, respectively. (c) A solution in the form of the
semi-structured DP beating stipe soliton. The parameters are (a) $s=1$ and $%
z=1.2$; (b) $s=-1$ and $z=0.9$; (c) $s=1$ and $z=1$, while maintaining $%
\protect\lambda _{R}=0$, $\protect\alpha =2$, $\protect\kappa =1$ and $%
\protect\Lambda $ satisfying condition~(\protect\ref{symmetry-dbs}) in all the
cases. The cyan curves represent the waveforms at $t=0$.}
\end{figure*}

The above asymptotic analysis demonstrates that each asymptotic soliton
emerges as a beating stripe structure formed by the bright-dark soliton
superposition, exhibiting spatiotemporal dual-periodic modulations governed
by $\Psi _{p}$ and $\Psi _{p}^{\prime }$ with the spatial and temporal
periods being $T_{x}=\pi /k_{m}$ and $T_{t}=4\pi /z^{2}$ (with $z$ mapped to
$\lambda $ as per Eq. (\ref{Jouk})), respectively. Furthermore, the opposite
signs in front of $\Psi _{p}$ and $\Psi _{p}^{\prime }$ with $j=1,2$ in Eq.~(%
\ref{asymptotic-p}) indicate the existence of out-of-phase configurations,
with respect to the two components, in system~(\ref{helicoidal SOC}). By
analyzing the $\Delta _{j}$ terms, one may derive trajectories and analyze
their slope, similar to Eqs.~(\ref{trajectory-bb}) and~(\ref{trajectory-bb-3}%
) for individual asymptotic solitons, which are not explicitly presented
here. The curved trajectories asymptotically approach the linear line $%
\mathcal{L}$, $\delta =-z(x+2\lambda _{R}t)=0$, with the symmetry condition
governing the soliton distribution about this asymptotic line given by
\begin{equation}
\begin{aligned} \label{symmetry-dbs} \Lambda^2S^4=z^2|z^2-s|. \end{aligned}
\end{equation}

Note that the system's total density, $|\mathbf{\Psi }|^{2}=\mathbf{\Psi }%
^{\dag }\mathbf{\Psi }=|\Psi _{1}|^{2}+|\Psi _{2}|^{2}$, preserves the
spatiotemporal aperiodicity, in spite of the intra-component periodicity.
The resultant DP solitons, in their dark/bright form with curved paths,\ are
revealed by inter-component superposition,
\begin{equation}
\begin{aligned} \label{total-dbs} \sum_{k=1}^2
|S_k^{j\pm}|^2=1+(S^2-1)\text{sech}^2\Delta, \end{aligned}
\end{equation}%
where $\Delta $ selects one element $\Delta _{j}$ ($j=1,2,3,4$) according to
the specific asymptotic soliton. This universal form bridges soliton types
with the same total density: $s=1$ generates bright MP states as shown in
Fig.~\ref{fig5}(a), while $s=-1$ produces their dark counterparts as shown
in Fig.~\ref{fig5}(b), demonstrating the identification of the soliton types
(bright/dark) through tunable interatomic interactions, attractive or
repulsive alike.

Consistent with the above asymptotic analysis, Fig.~\ref{fig5} illustrates a
static DP soliton exhibiting the simultaneous spatiotemporal periodicity
(manifested as beating and stripe patterns). The soliton's curved trajectory
asymptotically approaches the line $x=0$, while the total density
distribution remains nonperiodic. For the attractive interaction ($s=1$),
the total density features bright solitons [see Fig.~\ref{fig5}(a)], whereas
the repulsive interaction ($s=-1$) yields dark solitons, see Fig.~\ref{fig5}%
(b). The pseudospin components of these solitons exhibit out-of-phase
configurations. The spatial periodicity along the $x$-direction is governed
by the SOC strength $\alpha $ and helicity $\kappa $, while the temporal
periodicity is controlled by parameter $z$, which maps to the spectral
parameter $\lambda _{I}$ via the Joukowsky transformation (\ref{Jouk}). Note
that, for $s=1$ and $z=1$ (i.e., $\lambda _{I}=1$), a degenerate DP beating
stripe soliton with a left-handed structure emerges, see Fig.~\ref{fig5}(c);
setting $z=-1$ produces its right-handed degenerate counterpart. Such
degenerate structures are absent in the case of the repulsive interaction.

\begin{figure*}
\includegraphics[scale=0.45]{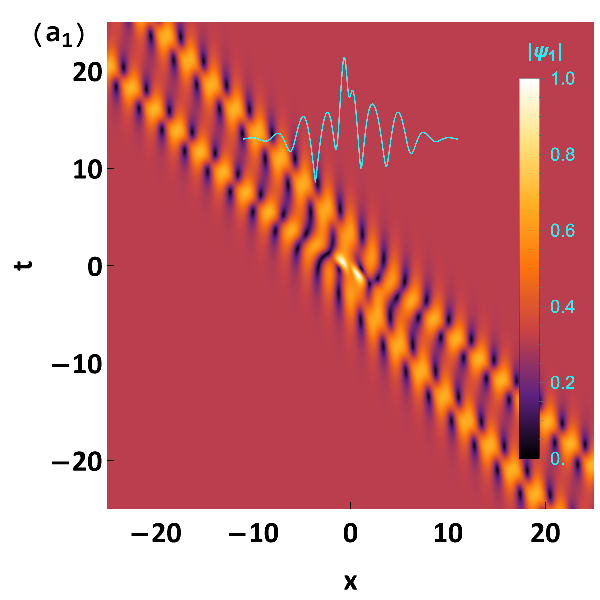}\hspace{3mm}
\includegraphics[scale=0.45]{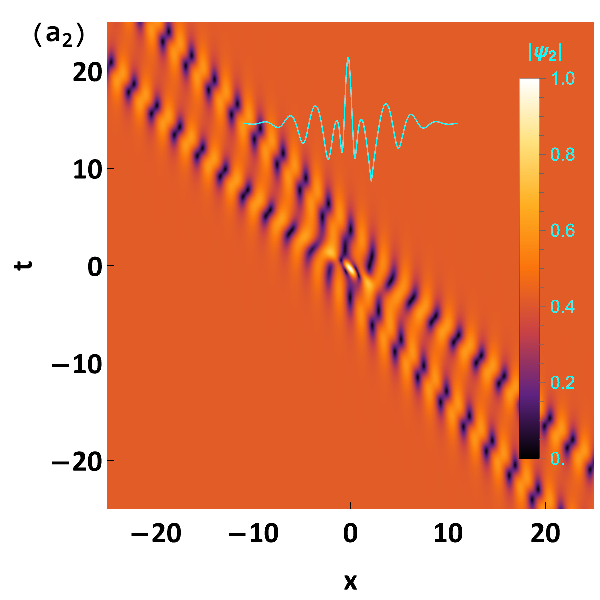} \hspace{3mm}
\includegraphics[scale=0.45]{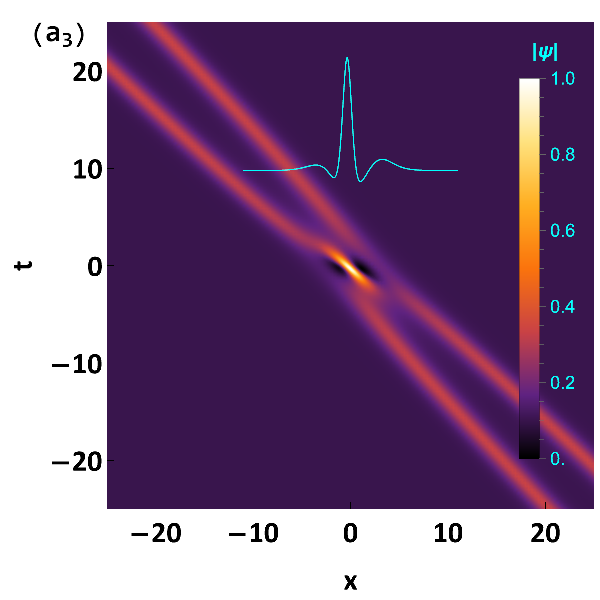} \hspace{3mm}\newline
\includegraphics[scale=0.45]{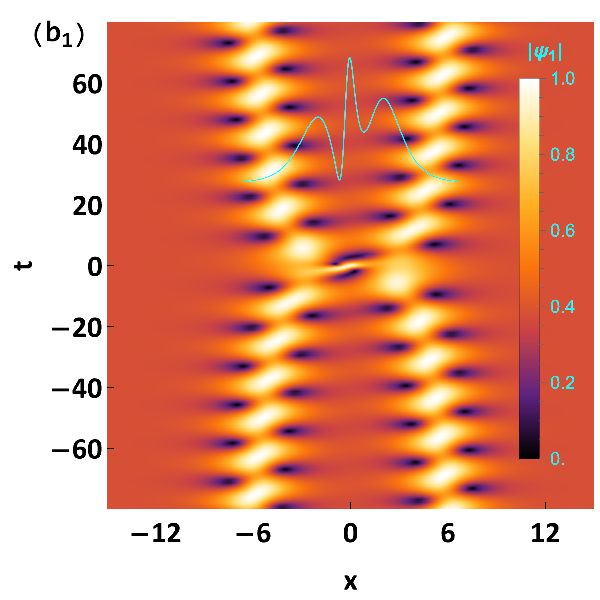}\hspace{3mm}
\includegraphics[scale=0.45]{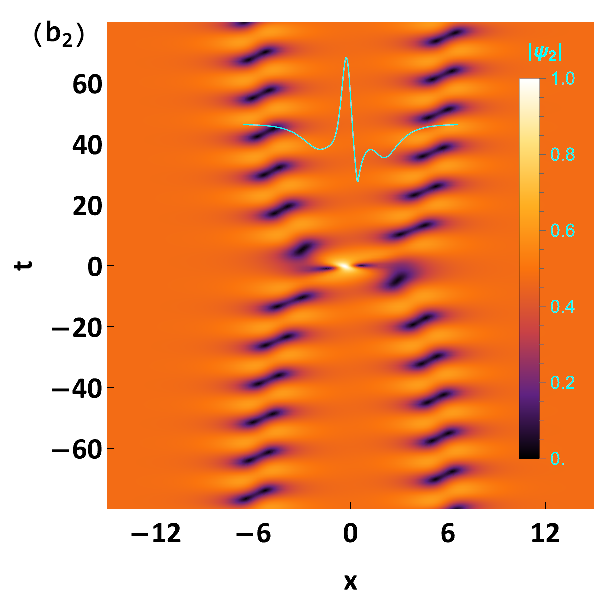} \hspace{3mm}
\includegraphics[scale=0.45]{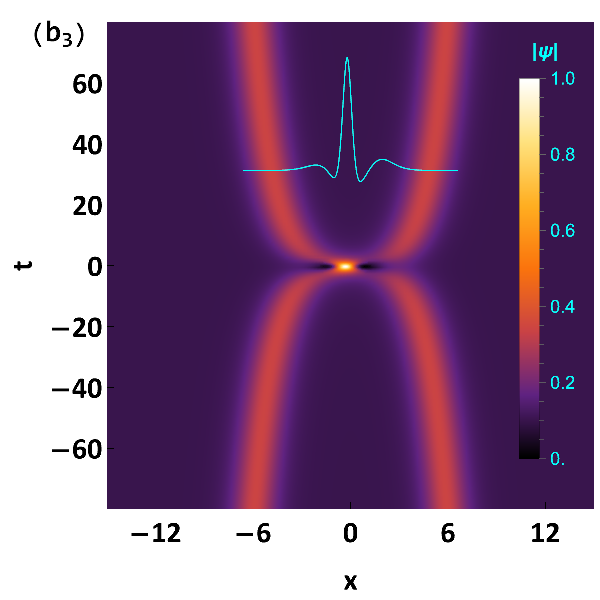} \hspace{3mm}\newline
\includegraphics[scale=0.45]{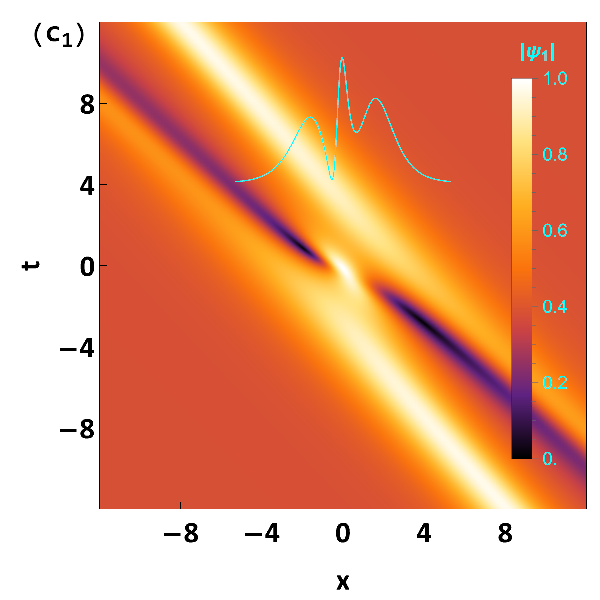}\hspace{3mm}
\includegraphics[scale=0.45]{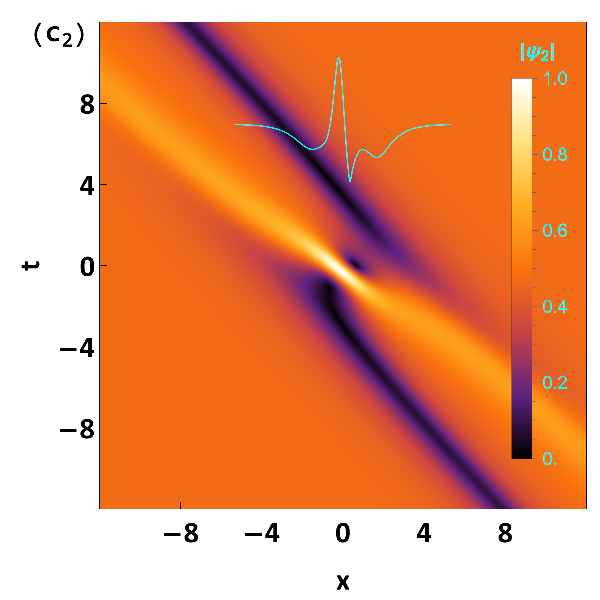} \hspace{3mm}
\includegraphics[scale=0.45]{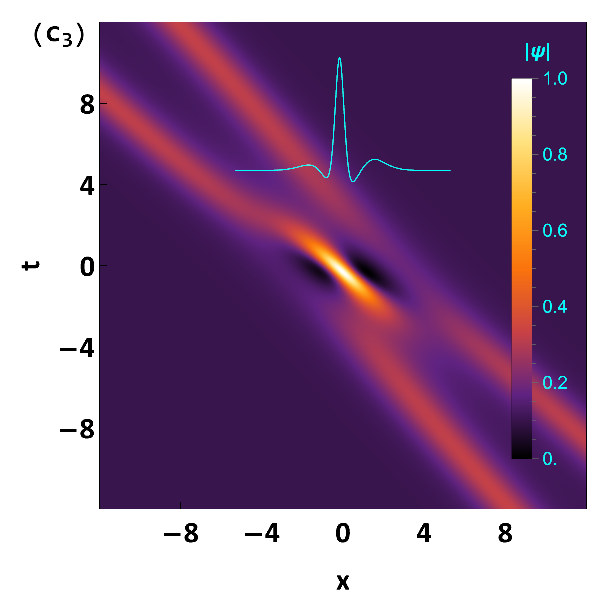} \hspace{3mm}\newline
\vspace{-3mm}
\caption{\label{fig6}Three degenerate structures of DP beating stripe solitons in case
of the attractive interaction ($s=1$), as produced by solutions~(\protect\ref%
{D-d-b2}). (a) A pure spatially-periodic state, with $\protect\alpha =1.5$, $%
\protect\kappa =0.5$ and $\protect\lambda _{R}=-\frac{\cos 2\protect\gamma }{%
4\cos \protect\gamma }$. (b) A pure temporally-periodic one, with $\protect%
\alpha =\frac{\protect\sqrt{3}}{8}$, $\protect\kappa =\frac{1}{8}$ and $%
\protect\lambda _{R}=-\frac{\cos \protect\gamma }{2}$. (c) A state which is
completely aperiodic in space and time, with $\protect\alpha =\frac{\protect%
\sqrt{3}}{8}$, $\protect\kappa =\frac{1}{8}$ and $\protect\lambda _{R}=-%
\frac{\cos 2\protect\gamma }{4\cos \protect\gamma }$. Other parameters are $%
\protect\gamma =\frac{\protect\pi }{3}$ and $\protect\Lambda ^{2}=\frac{\cos
\protect\gamma }{2}$. The cyan curves represent the waveforms at $t=0$.}
\end{figure*}

In the case of $s=1$ with $|\lambda _{I}|\leq 1$, the DP beating stripe
solitons exhibit several distinct structural configurations, according to
the solutions given by Eqs.~(\ref{D-d-b2}), as illustrated in Fig.~\ref{fig6}%
. These solitons, formed via the bright-dark superposition, demonstrate
spatiotemporal periodicity with spatial and temporal periods given by
\begin{equation}
\begin{aligned} \label{periods-dbs2} T_x=\frac{2\pi}{|2k_m-\cos\gamma|},~~
T_t=\frac{2\pi}{|4\lambda_R\cos\gamma+\cos2\gamma|}. \end{aligned}
\end{equation}%
The periodicity scales are tunable via parameters $\alpha $, $\kappa $ and $%
\gamma $ (related to $\lambda _{I}$ as per $\lambda _{I}=\sin \gamma $).
Thus, three degenerate cases can be obtained:

\begin{itemize}
\item[1] The temporal aperiodicity. When $\lambda _{R}=-\frac{\cos 2\gamma }{%
4\cos \gamma }$, the temporal period $T_{t}$ diverges, eliminating the
periodicity along $t$, as shown in Fig.~\ref{fig6}(a).

\item[2] The spatial aperiodicity. For $k_{m}=\frac{1}{2}\cos \gamma $, the
spatial period $T_{x}$ diverges, suppressing the $x$-periodicity, as shown
in Fig.~\ref{fig6}(b).

\item[3] The hybrid soliton. Simultaneously satisfying conditions $\lambda
_{R}=-\frac{\cos 2\gamma }{4\cos \gamma }$ and $k_{m}=\frac{\cos \gamma }{2}$%
, it yields a unique DP soliton with curved trajectories, combining dark and
bright branches in the out-of-phase configuration, with respect to the
pseudospin components $\Psi _{1,2}$, as shown in Fig.~\ref{fig6}(c).
\end{itemize}

Note that all the three degenerate structures retain nonperiodic total
density profiles characteristic of DP bright solitons. This parametric
control framework highlights the interplay between the SOC and spectral
parameter in engineering soliton hierarchies.

\begin{figure*}
\includegraphics[scale=0.45]{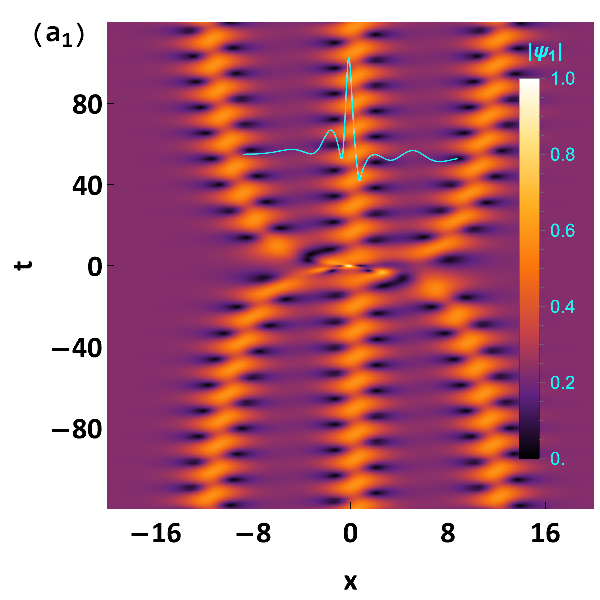}\hspace{3mm}
\includegraphics[scale=0.45]{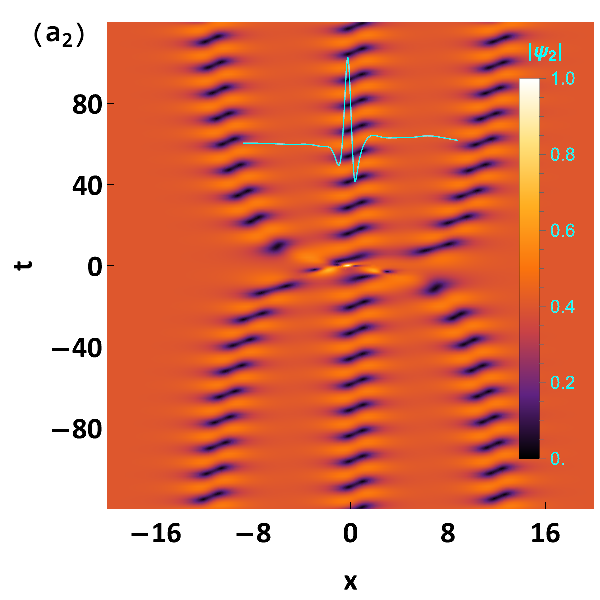} \hspace{3mm}
\includegraphics[scale=0.45]{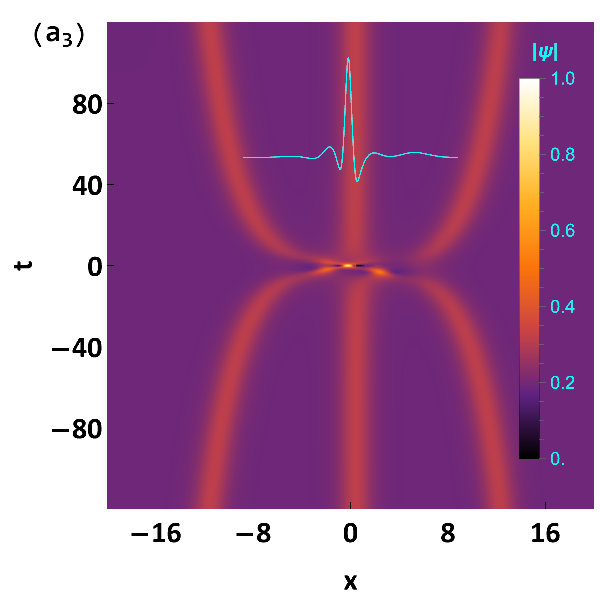} \hspace{3mm}\newline
\includegraphics[scale=0.45]{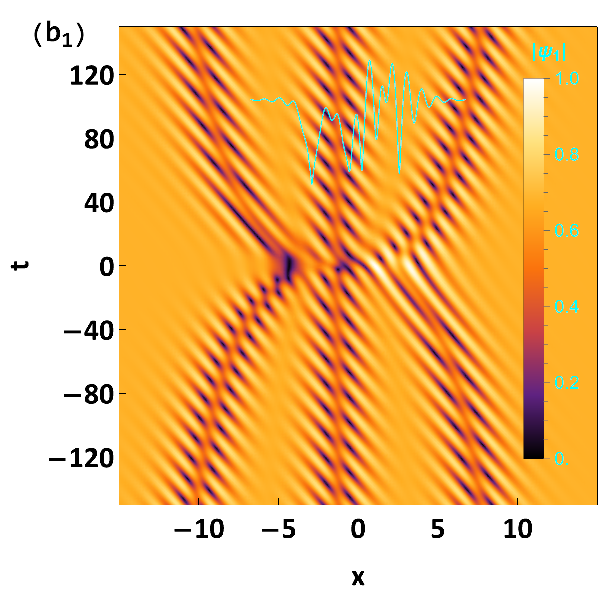}\hspace{3mm}
\includegraphics[scale=0.45]{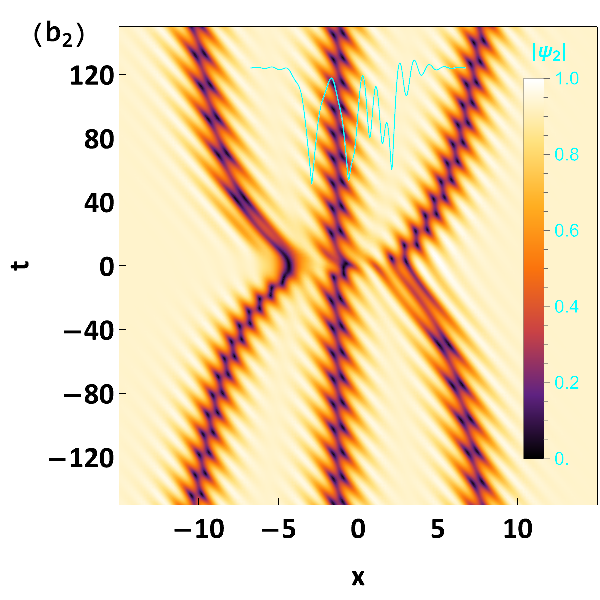} \hspace{3mm}
\includegraphics[scale=0.45]{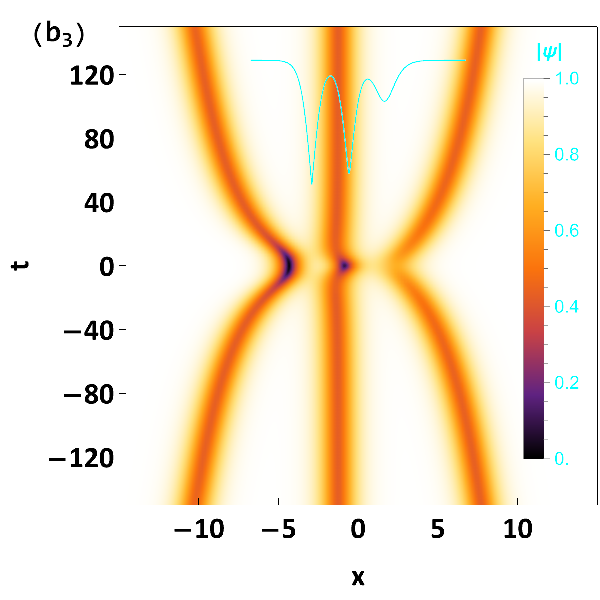} \hspace{3mm}\newline
\vspace{-3mm}
\caption{\label{fig7}Static triple-pole beating stipe solitons in components $|\Psi
_{1,2}|$ with the bright (a) and dark (b) total density $|\Psi |^{2}=|\Psi
_{1}|^{2}+|\Psi _{2}|^{2}$, in the case of the attractive and repulsive
interatomic interactions, respectively. The parameter sets are identical to
those used in Figs.~\protect\ref{fig6}(b) and \protect\ref{fig5}(b),
respectively. The cyan curves represent the waveforms at $t=0$.}
\end{figure*}

By further utilizing eigenfunction~(\ref{eigenfunction2}) and setting $n=1$,
$m_{k}=2$, we derive a triple-pole beating stripe soliton through the
general solution~(\ref{multi-pole}), as displayed in Fig.~\ref{fig7}. This
soliton retains the characteristic beating/stripe patterns and nonperiodic
total density (bright/dark) demonstrated above by the lower-order pole
solitons. However, distinct from the DP case presented above in Fig.~\ref%
{fig5}, it incorporates a central straight soliton flanked, on both sides,
by curved ones, with trajectories resembling the triple-pole bright solitons
in Fig.~\ref{fig3}(a).

For the triple-pole beating stripe solitons, parameter adjustments reproduce
the structures similar to those observed in Fig.~\ref{fig5} and the
degenerate configurations in Fig.~\ref{fig6}. Due to their structural
similarity (differing only in the trajectory geometry), they are not
replotted here. We stress that odd-order pole solitons (e.g., this
triple-pole ones) do not produce the half-structured solitons observed in
Fig.~\ref{fig5}(c), which exist solely in the even-order pole cases.

\section{Multi-pole beating stripe solitons and breathers on the periodic
background}

The above consideration dealt with two distinct seed configurations: (i)
dual zero-seed components, and (ii) one plane-wave seed paired with a
zero-seed component. We now extend the analysis to the case when both
initial seeds are plane waves, explicitly defined as $u_{j0}=a_{j}e^{i\theta
_{j}}$, where $\theta _{j}=k_{j}x+\left[ s(a_{1}^{2}+a_{2}^{2})-\frac{1}{2}%
k_{j}^{2}\right] t$. Here, $a_{j}$ and $k_{j}$ are real amplitudes and
wavenumbers of the seed solutions, respectively. The relative wavenumber $%
\delta =k_{1}-k_{2}$ plays a crucial role in governing the dynamics of the
resulting nonlinear localized waves. We therefore proceed by systematically
analyzing two regimes, \textit{viz}., the wavenumber-matched and mismatched
ones, which correspond to $\delta =0$ and $\delta \neq 0$, respectively,
each case exhibiting unique nonlinear phenomena.

\subsection{The wavenumber-matched case: $\protect\delta =0$}

In this configuration, by means of the matrix factorization, we obtain two
distinct eigenfunction solutions for the Lax pair~(\ref{A1}):
\begin{subequations}
\begin{align}
\boldsymbol{\Phi }_{1}(\lambda )=&
\begin{bmatrix}
e^{i\mathrm{A}(\tau _{1})} \\[3pt]
e^{i\theta _{1}}\left[ \frac{a_{1}e^{i\mathrm{A}(\tau _{1})}}{k_{1}+\tau _{1}%
}-a_{2}l_{3}e^{i\mathrm{A}(-k_{1})}\right] \\[8pt]
e^{i\theta _{1}}\left[ \frac{a_{2}e^{i\mathrm{A}(\tau _{1})}}{k_{1}+\tau _{1}%
}+a_{1}l_{3}e^{i\mathrm{A}(-k_{1})}\right]%
\end{bmatrix}%
,  \label{eg31} \\
\boldsymbol{\Phi }_{2}(\lambda )=&
\begin{bmatrix}
e^{i\mathrm{A}(\tau _{1})}+l_{3}e^{i\mathrm{A}(\tau _{2})} \\[3pt]
a_{1}e^{i\theta _{1}}\left[ \frac{e^{i\mathrm{A}(\tau _{1})}}{k_{1}+\tau _{1}%
}+\frac{l_{3}e^{i\mathrm{A}(\tau _{2})}}{k_{1}+\tau _{2}}\right] \\[8pt]
a_{2}e^{i\theta _{1}}\left[ \frac{e^{i\mathrm{A}(\tau _{1})}}{k_{1}+\tau _{1}%
}+\frac{l_{3}e^{i\mathrm{A}(\tau _{2})}}{k_{1}+\tau _{2}}\right]%
\end{bmatrix}%
.  \label{eg32}
\end{align}%
Here, $\mathrm{A}(\tau )=(\tau -\lambda )x+[\tau ^{2}/2-\lambda
^{2}-(a_{1}^{2}+a_{2}^{2})s]t$, $l_{3}$ is a nonzero constant, and $\tau
_{j} $ ($j=1,2$) are roots of the quadratic equation
\end{subequations}
\begin{equation}
\begin{aligned} \label{quadratic equationj}
\tau^2+(k_1-2\lambda)\tau-2k_1\lambda-s(a_1^2+a_2^2)=0. \end{aligned}
\end{equation}

\begin{figure*}
\includegraphics[scale=0.45]{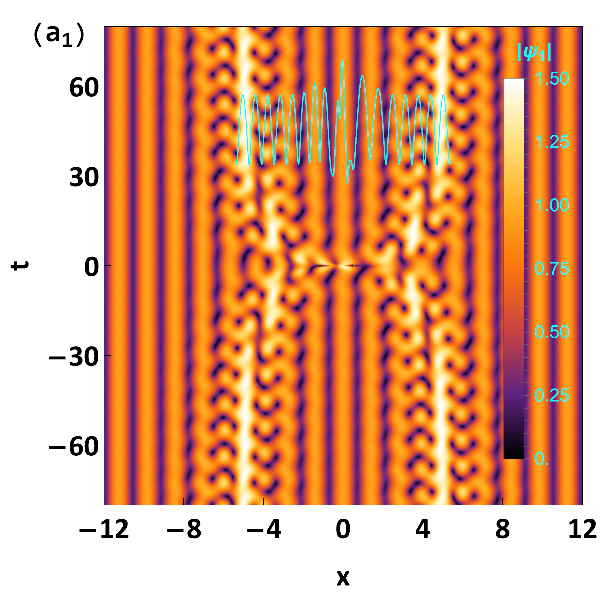}\hspace{3mm}
\includegraphics[scale=0.45]{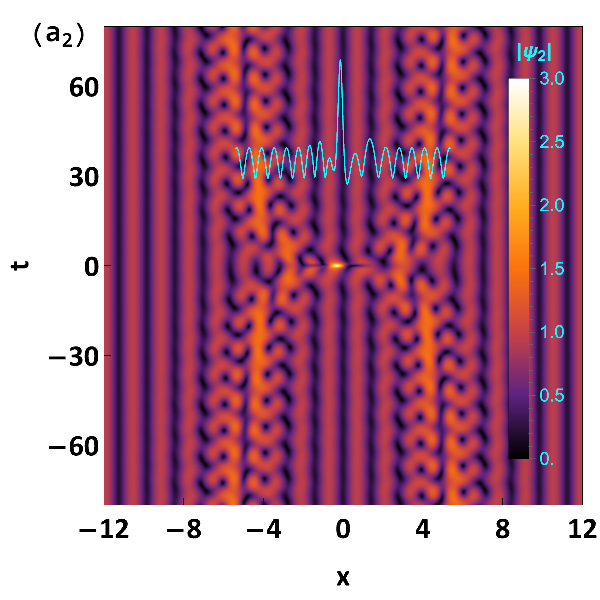} \hspace{3mm}
\includegraphics[scale=0.45]{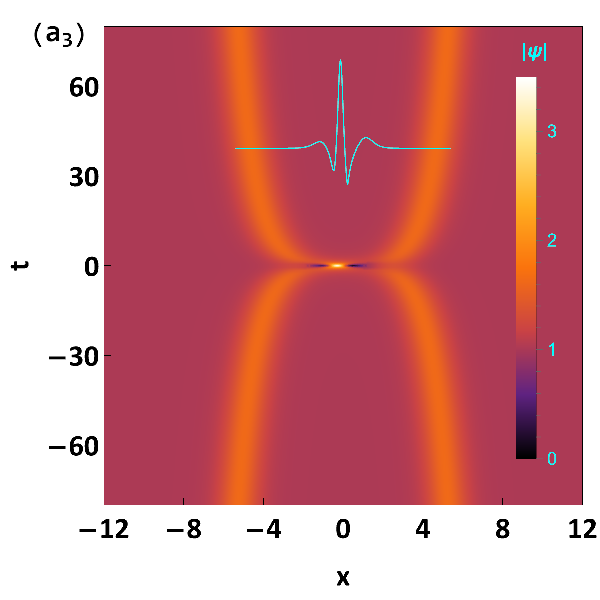} \hspace{3mm}\newline
\includegraphics[scale=0.45]{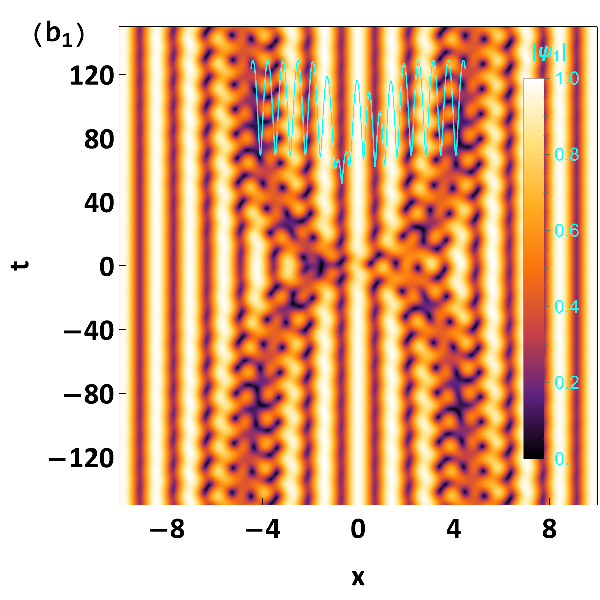}\hspace{3mm}
\includegraphics[scale=0.45]{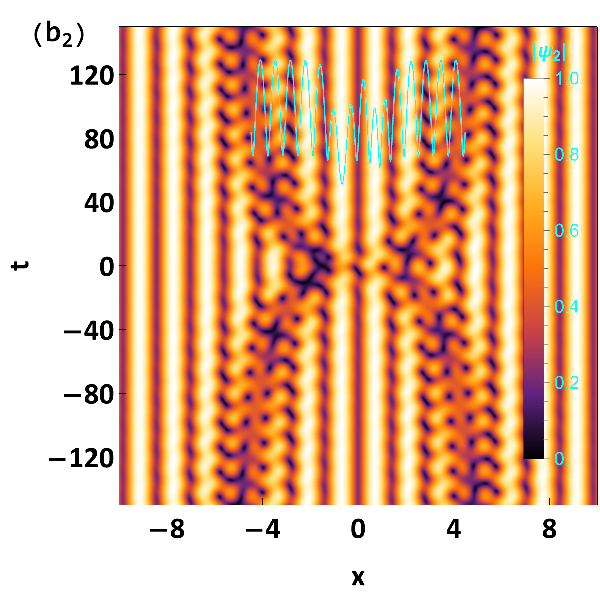} \hspace{3mm}
\includegraphics[scale=0.45]{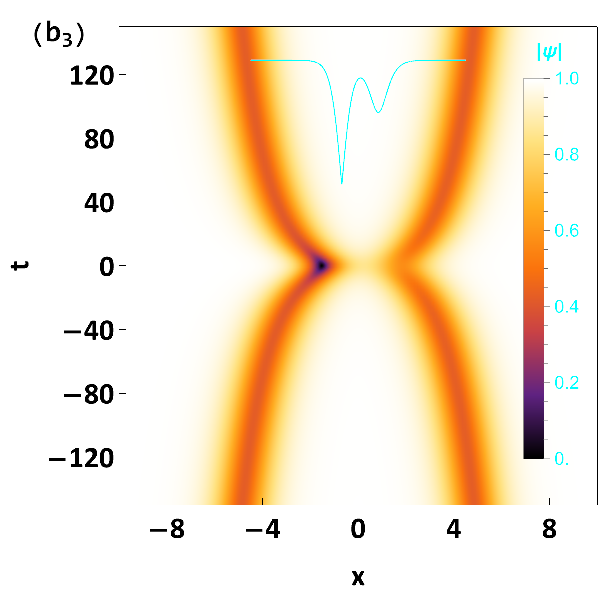} \hspace{3mm}\newline
\includegraphics[scale=0.45]{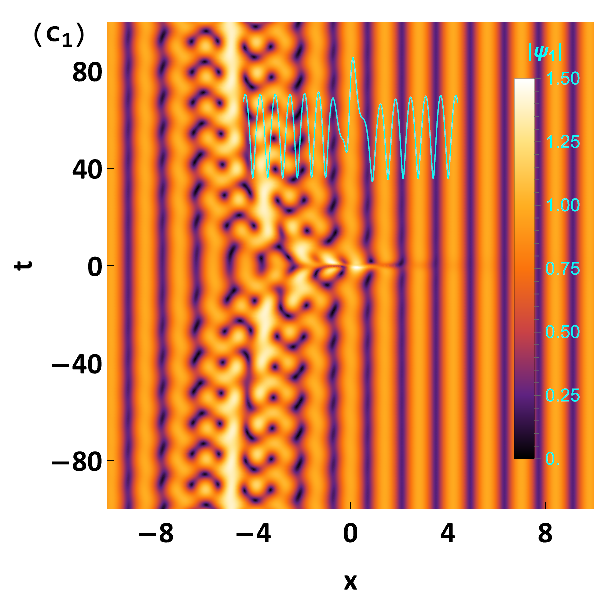}\hspace{3mm}
\includegraphics[scale=0.45]{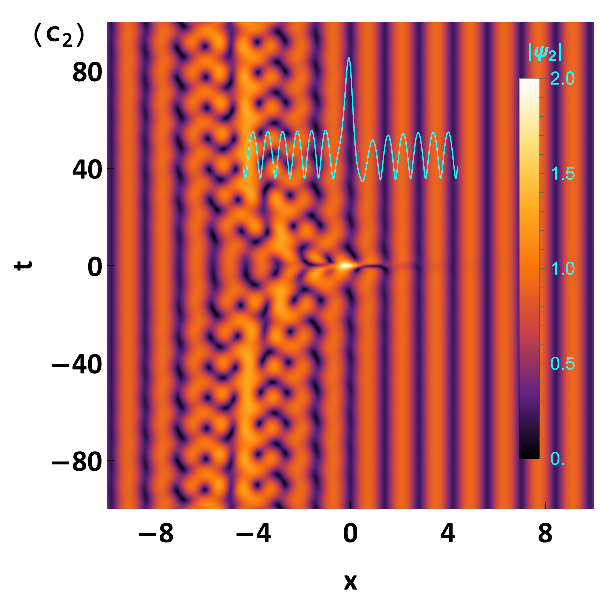} \hspace{3mm}
\includegraphics[scale=0.45]{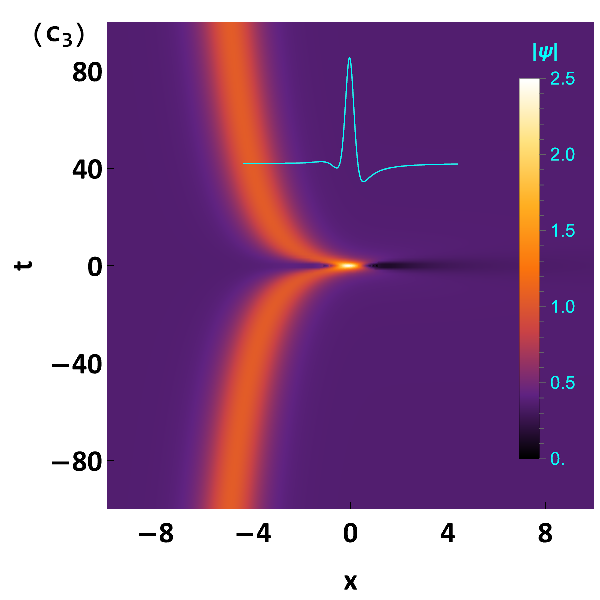} \hspace{3mm}\newline
\vspace{-3mm}
\caption{\label{fig8}Static DP beating stipe solitons on the periodic background in
components $|\Psi _{1,2}|$ with the bright (a) and dark (b) total density, $%
|\Psi |^{2}=|\Psi _{1}|^{2}+|\Psi _{2}|^{2}$, in the cases of the attractive
and repulsive interatomic interactions, respectively. (c) A solution in the
form of the semi-structured DP beating stipe soliton. Parameters are the
same as in Fig.~\protect\ref{fig5}, except for $a_{1,2}=1/\protect\sqrt{2}$.
The cyan curves represent the waveforms at $t=0$.}
\end{figure*}

Substituting eigenfunctions~(\ref{eg31}) and~(\ref{eg32}) into general
solutions~(\ref{multi-pole}), we construct two classes of localized wave
solutions on top of the periodic backgrounds. First, using eigenfunction~(%
\ref{eg31}), we derive MP beating stripe solitons formed by the bright-dark
soliton superposition. Unlike the case addressed in Section IV, the presence
of the dual plane-wave seeds introduces periodic backgrounds via gauge
transformation~(\ref{trans1}), as shown in Fig.~\ref{fig8}.

With the relative wavenumber $\delta =k_{1}-k_{2}=0$ (i.e., $\theta
_{1}=\theta _{2}$), the periodic background exhibits spatial periodicity
through transformation~(\ref{trans1}). The periodic backgrounds are
characterized by
\begin{equation}
\begin{aligned} \label{per-bg} |\Psi_1|^2_{{\rm
bg}}=&a_1^2\nu_+^2+a_2^2\nu_-^2+2a_1a_2\nu_+\nu_-\cos(2k_mx),\\
|\Psi_2|^2_{{\rm
bg}}=&a_2^2\nu_+^2+a_1^2\nu_-^2-2a_1a_2\nu_+\nu_-\cos(2k_mx), \end{aligned}
\end{equation}%
where the background period is $T_{\mathrm{bg}}=\pi /k_{m}$. Note that the
periodic backgrounds in both components are in the perfect out-of-phase
state. The out-of-phase relation of the components extends to the MP beating
stripe solitons. The superposition of the components yields a total density
profile that is a generic MP soliton without periodicity (both in its
background and soliton structures), as demonstrated in the right column of
Fig.~\ref{fig8}.

The beating stripe solitons which are formed by the bright-dark
superposition interact with the periodic background, generating intricate
periodic patterns. In the attractive ($s=1$) and repulsive ($s=-1$) cases,
the total density manifests as MP bright or dark solitons, respectively. We
stress that, in the case of attraction with spectral parameter $\lambda
_{I}=\pm (a_{1}^{2}+a_{2}^{2})$, the degenerate DP beating stripe solitons
on the periodic backgrounds feature the half-structured shape, \textit{viz}%
., the left- and right-half structures at $\lambda
_{I}=+(a_{1}^{2}+a_{2}^{2})$ (Fig.~\ref{fig8}(c)) and $\lambda
_{I}=-(a_{1}^{2}+a_{2}^{2})$, respectively.

\begin{figure*}
\includegraphics[scale=0.45]{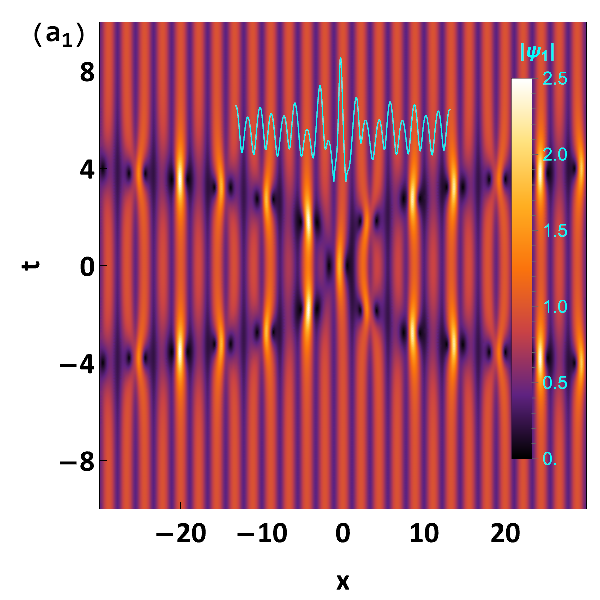}\hspace{3mm}
\includegraphics[scale=0.45]{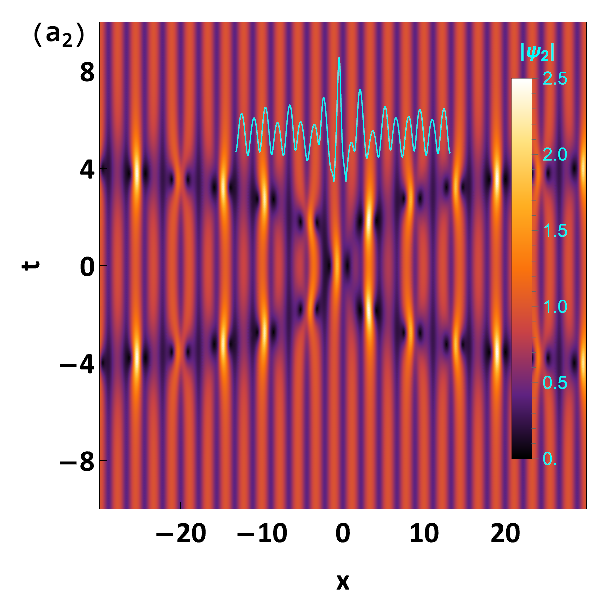} \hspace{3mm}
\includegraphics[scale=0.45]{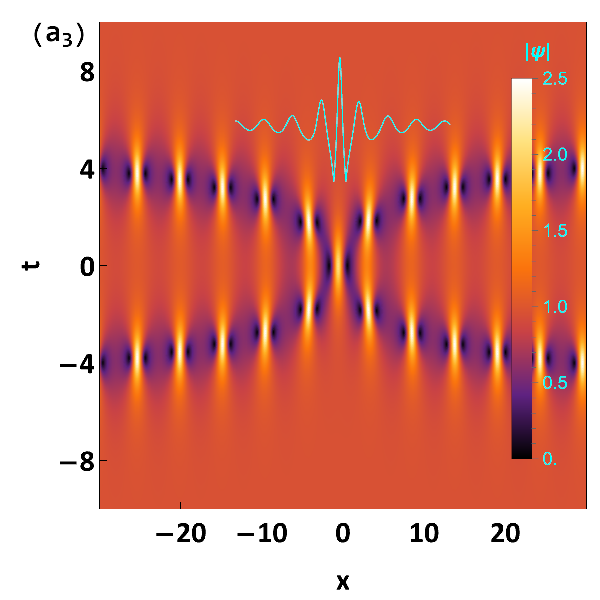} \hspace{3mm}\newline
\includegraphics[scale=0.45]{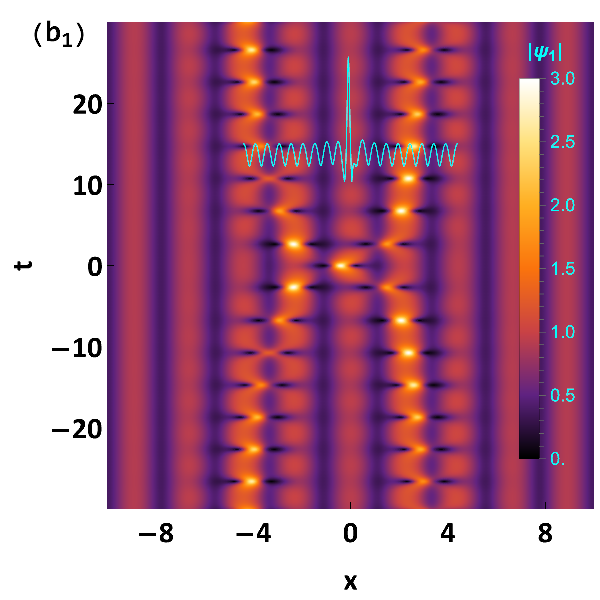}\hspace{3mm}
\includegraphics[scale=0.45]{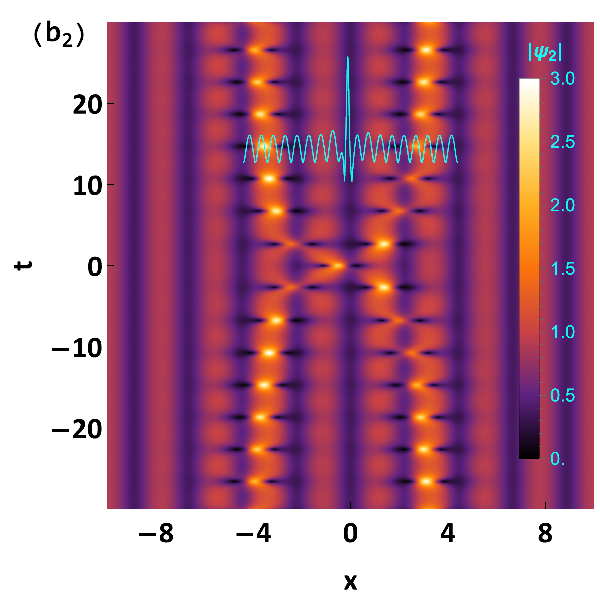} \hspace{3mm}
\includegraphics[scale=0.45]{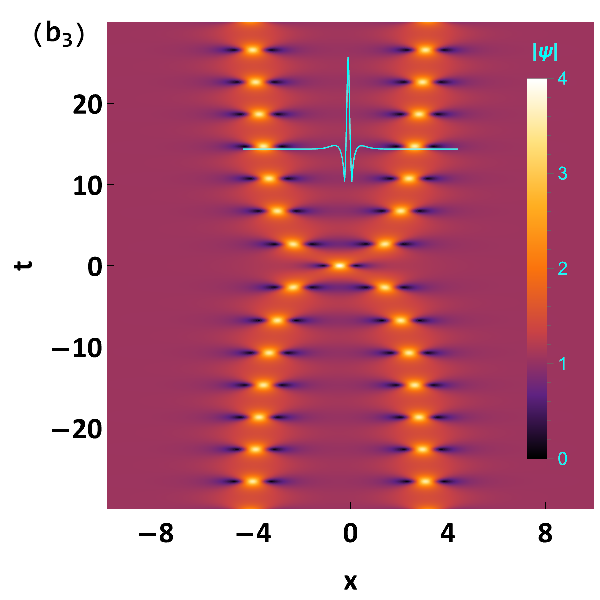} \hspace{3mm}\newline
\vspace{-3mm}
\caption{\label{fig9}Two kinds of DP breathers on top of the spatial periodic
background. The parameters are $\protect\lambda =0.8i$ for (a) and $\protect%
\lambda =1.2i$ for (b), with other parameters fixed as $s=1$, $a_{1}=a_{2}=1/%
\protect\sqrt{2}$, $l_{3}=1$, $k_{1}=k_{2}=0$ and $\protect\alpha =\protect%
\kappa =1$. The cyan curves represent the waveforms at $t=0$.}
\end{figure*}

By substituting the second eigenfunction~(\ref{eg32}) into general
solutions~(\ref{multi-pole}), we obtain DP breathers on periodic
backgrounds, as shown in Fig.~\ref{fig9}. Similarly, due to the vanishing
relative wavenumber ($\delta =0$), the periodic background, produced by
transformation~(\ref{trans1}), exhibits only spatial periodicity governed by
Eq.~(\ref{per-bg}). Despite out-of-phase periodic backgrounds, the
in-phase breather superposition cancels the background periodicity, leaving
plane-wave-based DP breathers in the total density profile.

The SOC strength $\alpha $ and helicity $\kappa $ modulate the period and
amplitude of the periodic background, whereas the breather's spatiotemporal
periodicity, structure, and amplitude are controlled by spectral parameter $%
\lambda $. Specifically, condition $0<|\lambda _{I}|<(a_{1}^{2}+a_{2}^{2})$
produces DP breathers which are asymptotically parallel to the $x$-axis, as
shown in Fig.~\ref{fig9}(a), while conditions $|\lambda
_{I}|>(a_{1}^{2}+a_{2}^{2})$ and $k_{1}=0$ produce the DP breathers which
are asymptotically parallel to the $t$-axis, as shown in Fig.~\ref{fig9}(b).
Breathers are oriented obliquely with respect to both axes when neither
condition is met. Note that, in the limit of $\lambda \rightarrow
(a_{1}^{2}+a_{2}^{2})i$, the breather's period diverges, causing
degeneration of the solution into a rogue wave on the periodic background.

\subsection{The wavenumber-mismatched case: $\protect\delta \neq 0$}

In this case, unequal wavenumbers lead to different initial phases for the
two components, positioning the MP breathers on top of a spatiotemporal
periodic background. Using the matrix decomposition method outlined above,
we derive the respective eigenfunction solutions for the Lax pair~(\ref{A1}%
):
\begin{equation}
\begin{aligned} \label{eigen31} \boldsymbol{\Phi}(\lambda)=& \begin{bmatrix}
e^{-i\mathrm{A}(\tau_1)}+e^{-i\mathrm{A}(\tau_2)} \\[3pt]
a_1e^{i\theta_1}\left[\frac{e^{-i\mathrm{A}(\tau_1)}}{k_1-\tau_1}
+\frac{e^{-i\mathrm{A}(\tau_2)}}{k_1-\tau_2}\right]\\[8pt]
a_2e^{i\theta_2}\left[\frac{e^{-i\mathrm{A}(\tau_1)}}{k_2-\tau_1}
+\frac{e^{-i\mathrm{A}(\tau_2)}}{k_2-\tau_2}\right] \end{bmatrix}.
\end{aligned}
\end{equation}%
Here, $\mathrm{A}(\tau )=(\tau +\lambda )x+[\lambda
^{2}+(a_{1}^{2}+a_{2}^{2})s-\tau ^{2}/2]t$, and $\tau _{j}$ ($j=1,2$) are
any two of the three distinct roots of the cubic equation
\begin{equation}
\begin{aligned} \label{cubic equationj}
\tau^3+&(2\lambda-k_1-k_2)\tau^2+[k_1k_2-2(k_1+k_2)\lambda-(a_1^2+a_2^2)s]%
\tau\\ &+(k_2a_1^2+k_1a_2^2)s+2k_1k_2\lambda=0. \end{aligned}
\end{equation}

Substituting the above eigenfunction solution into Eq.~(\ref{multi-pole})
makes it possible to construct MP breathers on top of the spatiotemporal
periodic background. Because the relative wavenumber is nonzero, the
periodic background differs from the case when the wavenumber vanishes. It
exhibits periodicity in both space and time, described by
\begin{equation}
\begin{aligned} \label{per-bg2} |\Psi_1|^2_{{\rm
bg}}=&a_1^2\nu_+^2+a_2^2\nu_-^2+2a_1a_2\nu_+\nu_-|\omega_{\pm}|
\cos(\varphi+2\arg\omega_{\pm}),\\ |\Psi_2|^2_{{\rm
bg}}=&a_2^2\nu_+^2+a_1^2\nu_-^2-2a_1a_2\nu_+\nu_-|\omega_{\pm}|
\cos(\varphi+2\arg\omega_{\pm}), \end{aligned}
\end{equation}%
where $\varphi =\frac{1}{2}\delta \lbrack 2x-(k_{1}+k_{2})t]-2k_{m}x$ and
\begin{equation}
\begin{aligned} \omega_{\pm}=\frac{(k_2-\tau_{1,2})(k_1-\tau_{1,2}^*)}
{(k_1-\tau_{1,2})(k_2-\tau_{1,2}^*)} \end{aligned}
\end{equation}%
with $-$ and $+$ in $\pm $ referring to the states before and after the
interaction between the periodic background and MP breathers, respectively.

\begin{figure*}
\includegraphics[scale=0.45]{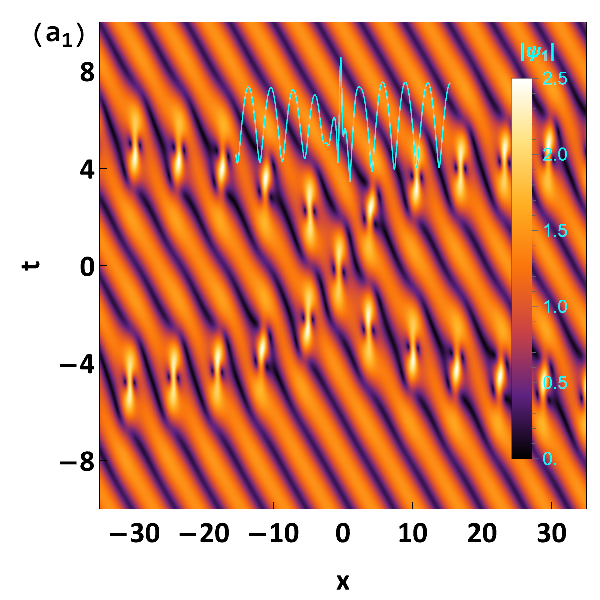}\hspace{3mm}
\includegraphics[scale=0.45]{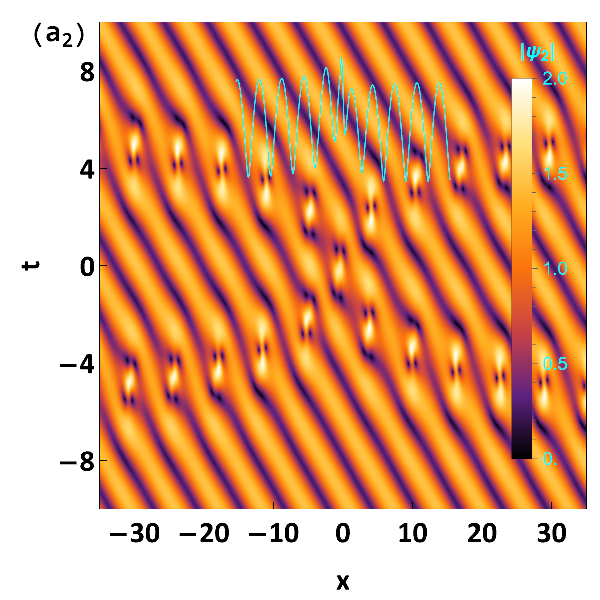} \hspace{3mm}
\includegraphics[scale=0.45]{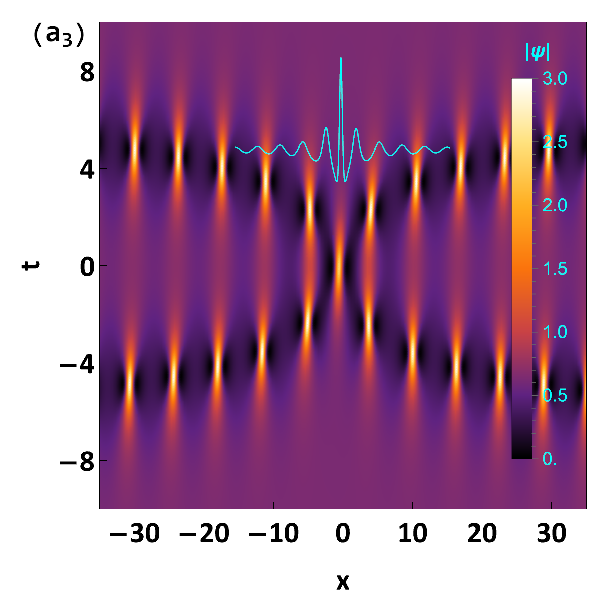} \hspace{3mm}\newline
\includegraphics[scale=0.45]{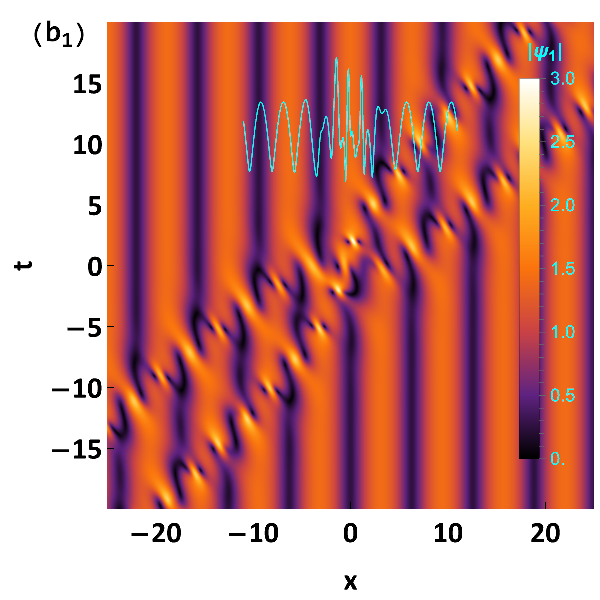}\hspace{3mm}
\includegraphics[scale=0.45]{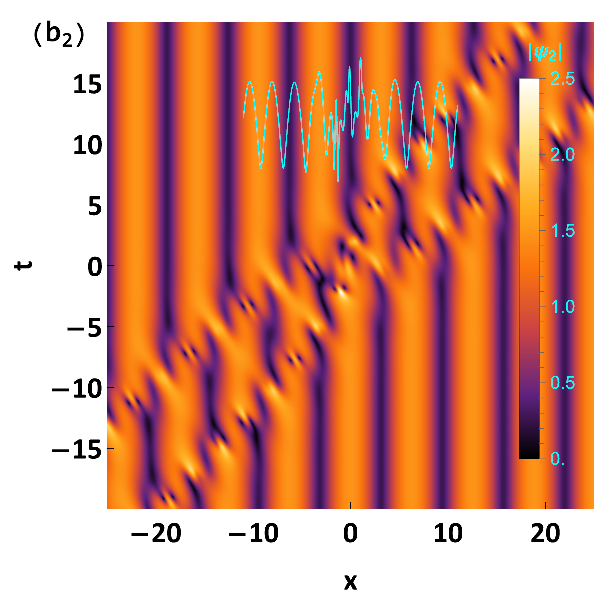} \hspace{3mm}
\includegraphics[scale=0.45]{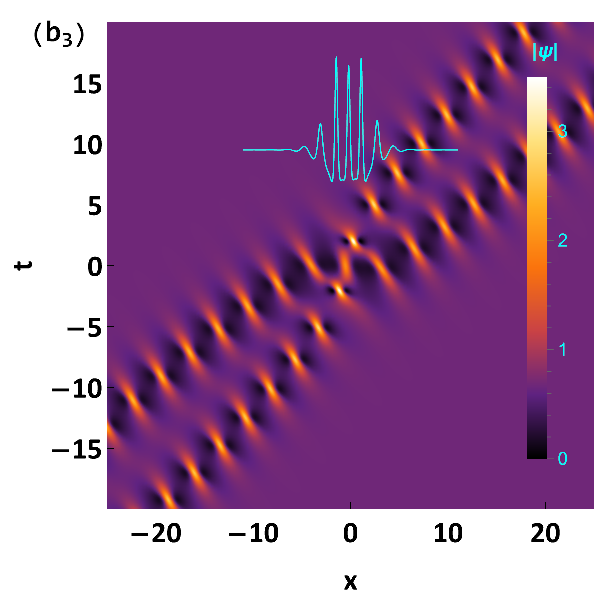} \hspace{3mm}\newline
\vspace{-3mm}
\caption{\label{fig11}Two kinds of DP breathers constructed on top of the spatiotemporal
periodic background. The parameters are $\protect\lambda =\protect\sqrt{(-5+2%
\protect\sqrt{2}i)/6}-0.5$, $k_{1}=2$, $k_{2}=0$ for (a) and $k_{1}=1$, $%
k_{2}=-1$, $\protect\lambda =i$ for (b), with other parameters fixed as $s=1$%
, $a_{1}=a_{2}=1$, $\protect\alpha =\protect\sqrt{2}$ and $\protect\kappa %
=1/2$. The cyan curves represent the waveforms at $t=0$.}
\end{figure*}

The spatiotemporal periodic background exhibits spatial and temporal periods
$T_{x}=2\pi /|\delta -2k_{m}|$ and $T_{t}=4\pi /|k_{1}^{2}-k_{2}^{2}|$,
respectively. Note that the two components maintain out-of-phase
configurations, with the periodic background preserving its periodicity
after the interaction with the MP breathers, while acquiring identical phase
shifts in both components. This ensures the persistent out-of-phase
relation, as confirmed by Fig.~\ref{fig11}. Further analysis of MP breathers
reveals that, when $\tau _{1I}=\tau _{2I}$, the asymptotic trajectories of
the DP breathers are parallel to the $x$-axis, as shown in Fig.~\ref{fig11}%
(a), while, under the condition of $\tau _{1R}\tau _{1I}=\tau _{2R}\tau _{2I}
$, the asymptotic trajectories are parallel to the $t$-axis. Generic DP
breathers can be produced when neither condition holds. The periodicity of
the background is tunable via the SOC parameters $(\alpha ,\kappa )$ and
wavenumbers $(k_{1},k_{2})$. Adjusting these parameters, one can produce the
time-only periodic background (infinite $T_{x}$, with $\delta =2k_{m}$) and
space-only periodic one (infinite $T_{t}$, with $k_{1}=-k_{2}$), as shown in
Fig.~\ref{fig11}(b).

Although complex wave structures arise with the periods which are
independently adjustable by means of SOC, spectral, and wavenumber
parameters, the total density always reduces to a standard MP breather. This
occurs because the breathers share identical periods while the background
components remain out-of-phase, causing mutual cancellation of their
periodic modulations in the total density profile.

\section{Numerical simulations}

\begin{figure}[ht]
\includegraphics[scale=0.27]{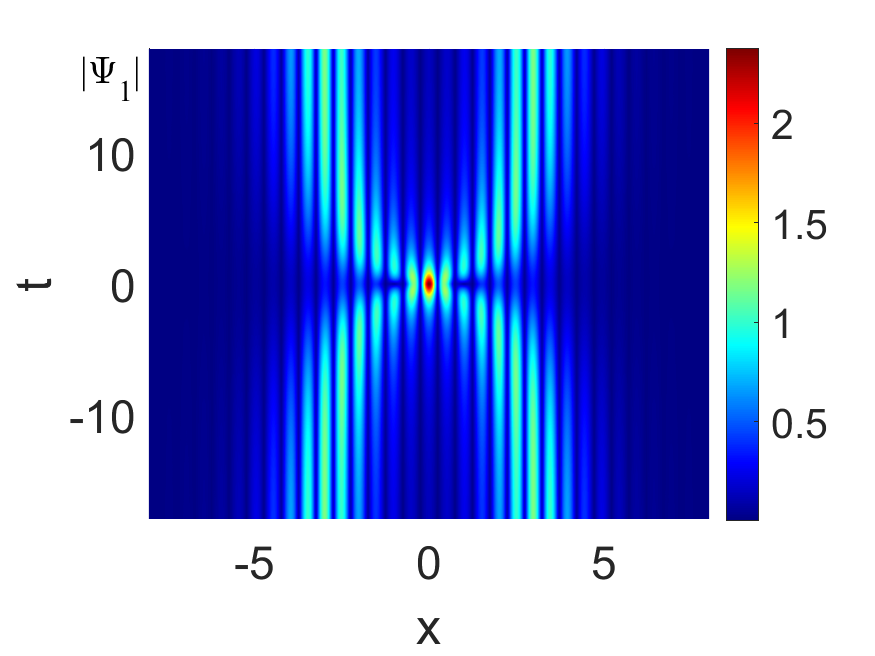}\hspace{3mm} %
\includegraphics[scale=0.27]{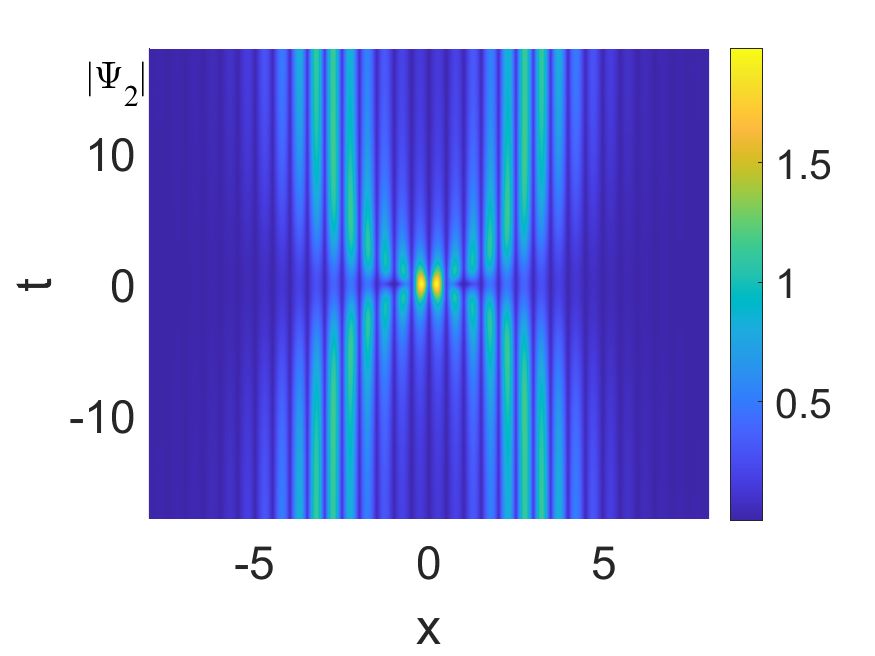}\newline
\includegraphics[scale=0.27]{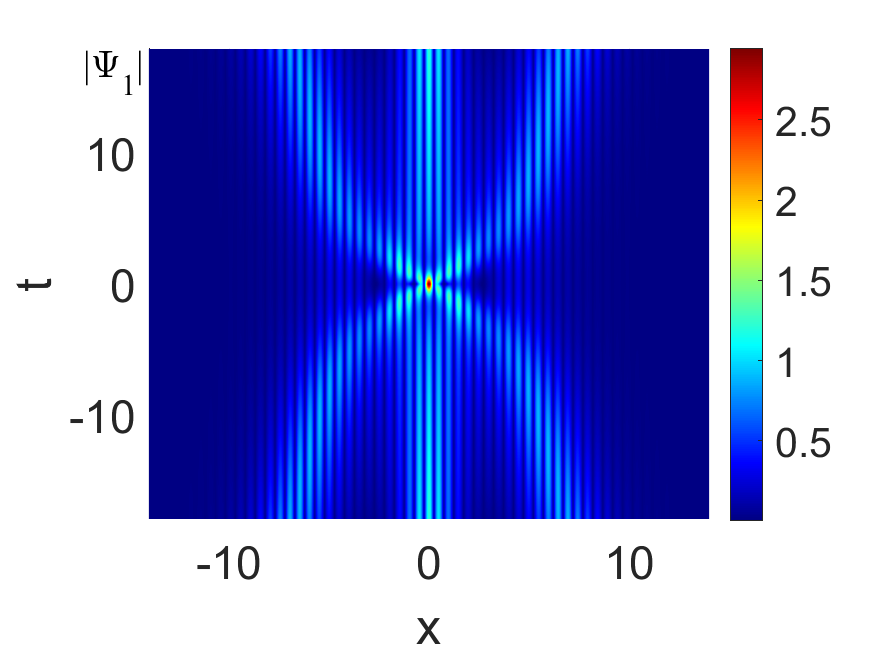}\hspace{3mm} %
\includegraphics[scale=0.27]{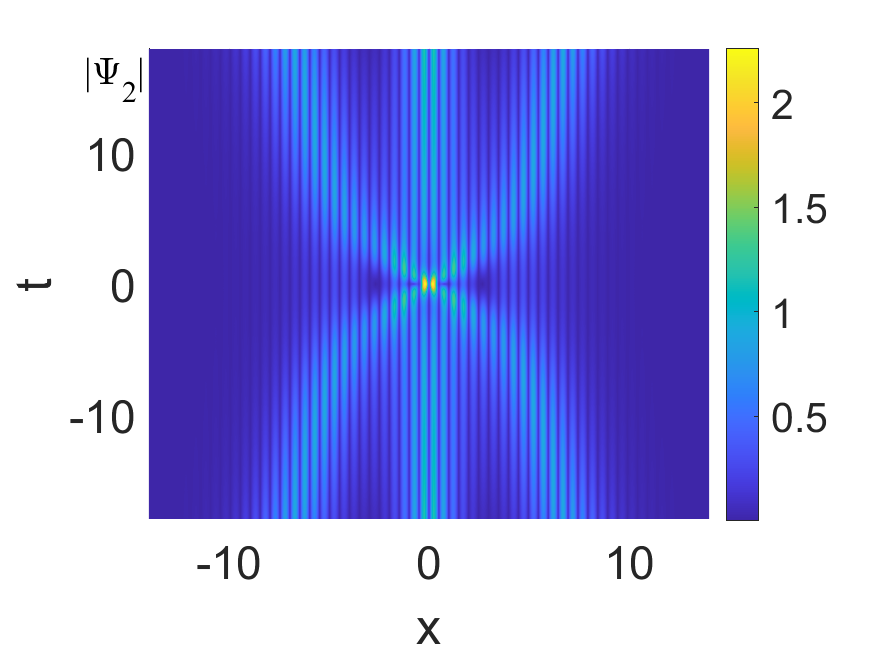}\newline
\vspace{-3mm}
\caption{Numerical simulations demonstrating the stable perturbed evolution
of the DP (top) and triple-pole (bottom) stripe solitons, whose unperturbed
forms is displayed in Figs.~\protect\ref{fig1}(a) and \protect\ref{fig3}(a),
respectively, with the addition of random perturbations at the $2\%$ level.}
\label{fig12}
\end{figure}

\begin{figure}[th]
\includegraphics[scale=0.27]{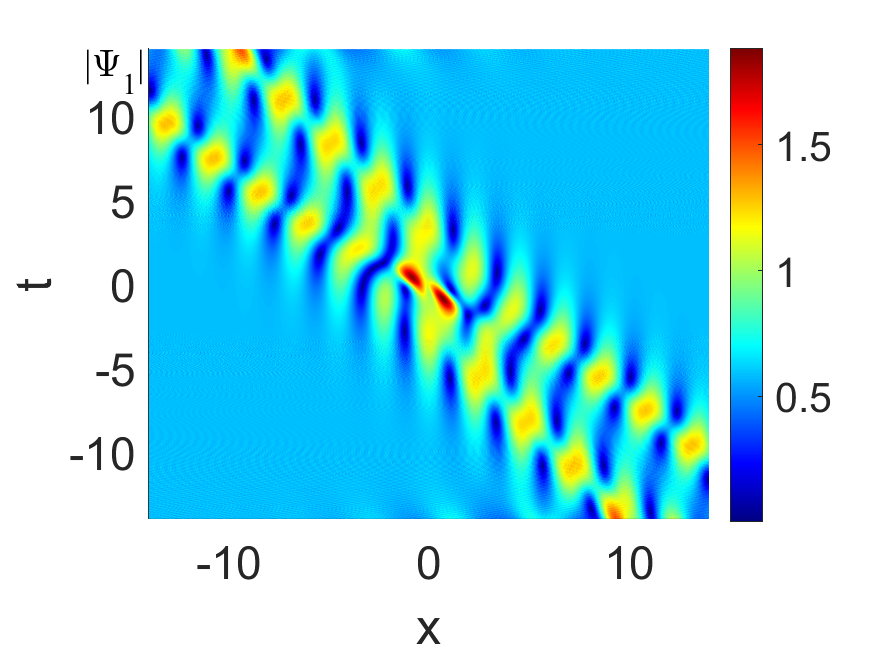}\hspace{3mm} %
\includegraphics[scale=0.27]{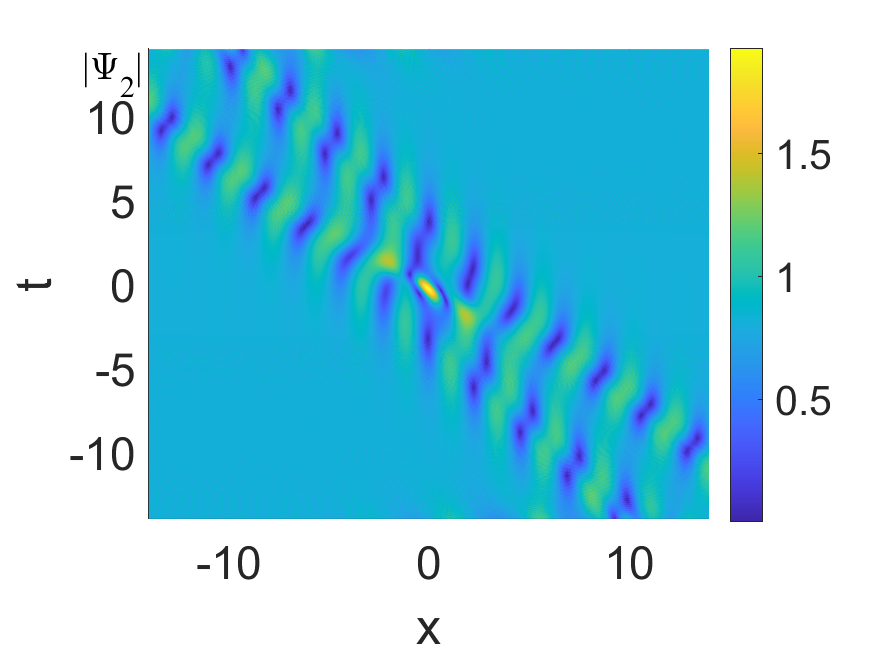}\newline
\includegraphics[scale=0.27]{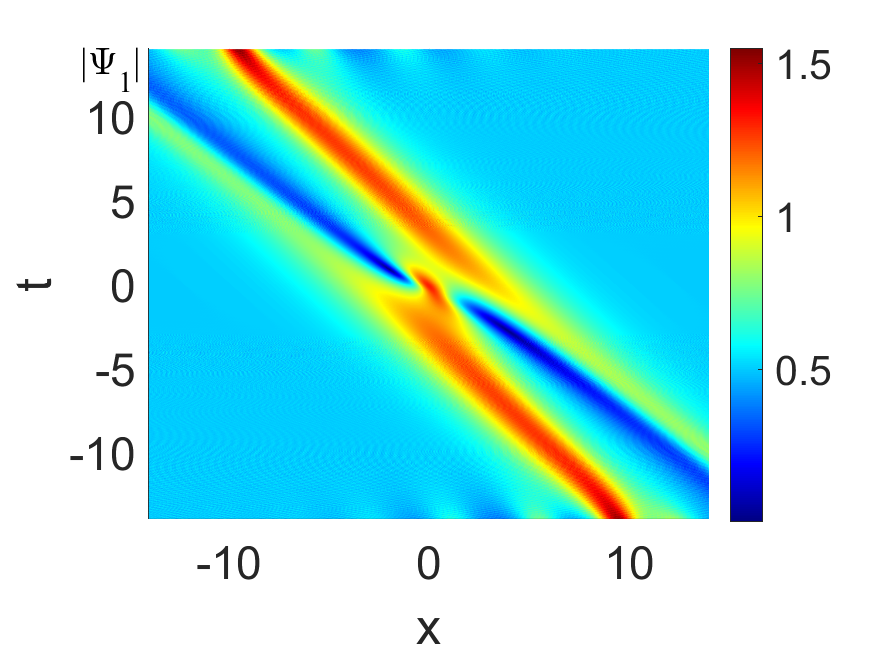}\hspace{3mm} %
\includegraphics[scale=0.27]{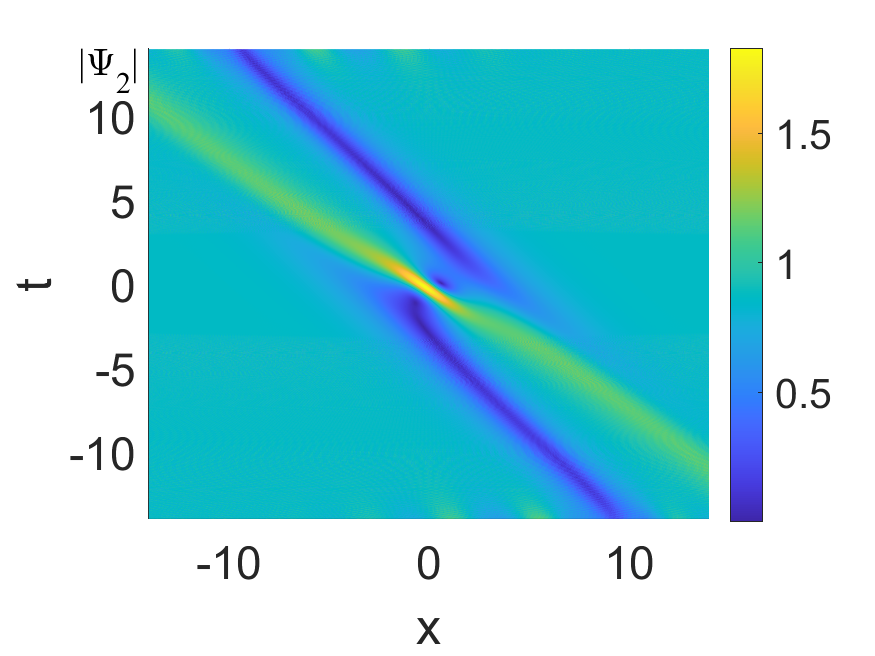}\newline
\vspace{-3mm}
\caption{Numerical simulations of the DP beating stripe solitons (top) and
its degenerate form (bottom), from Figs.~\protect\ref{fig6}(a) and \protect
\ref{fig6}(c), respectively with the addition od random perturbations at the
$2\%$ level. }
\label{fig13}
\end{figure}

\begin{figure}[th]
\includegraphics[scale=0.27]{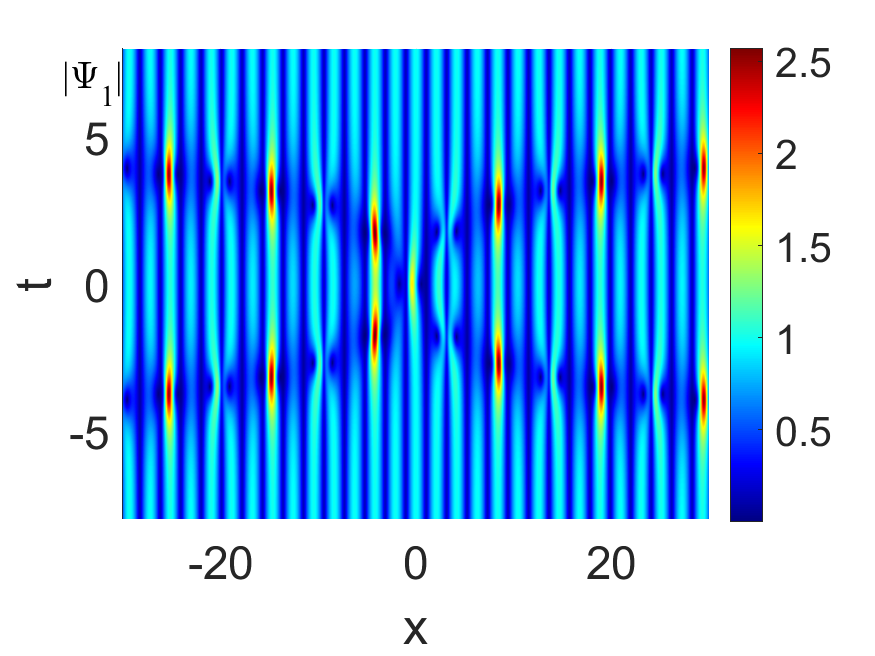}\hspace{3mm} %
\includegraphics[scale=0.27]{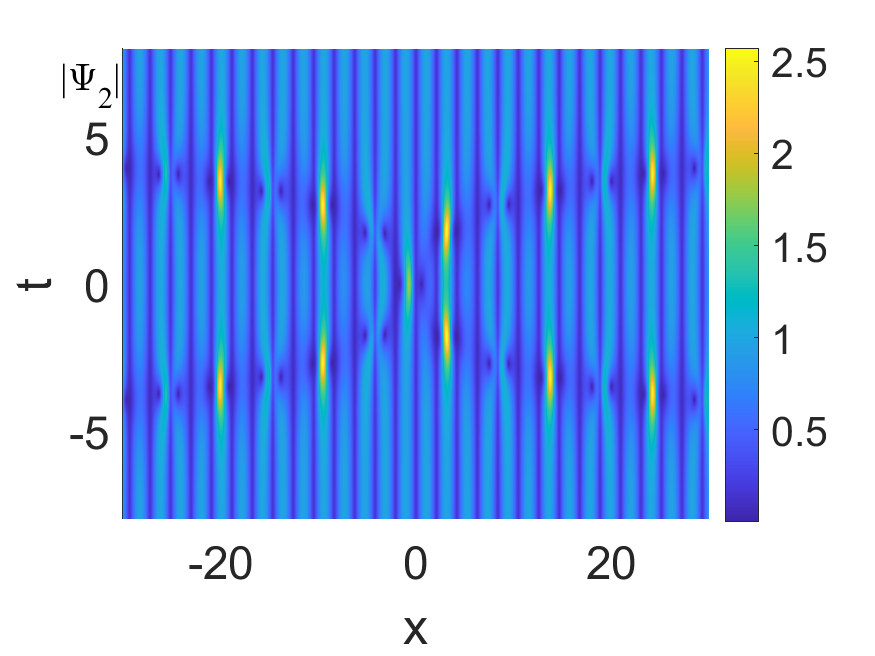}\newline
\includegraphics[scale=0.27]{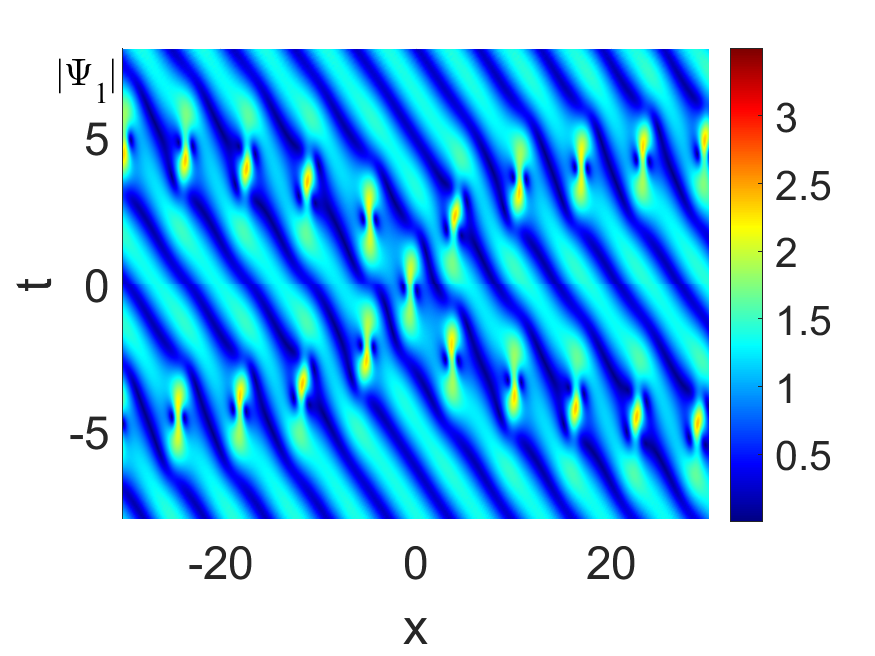}\hspace{3mm} %
\includegraphics[scale=0.27]{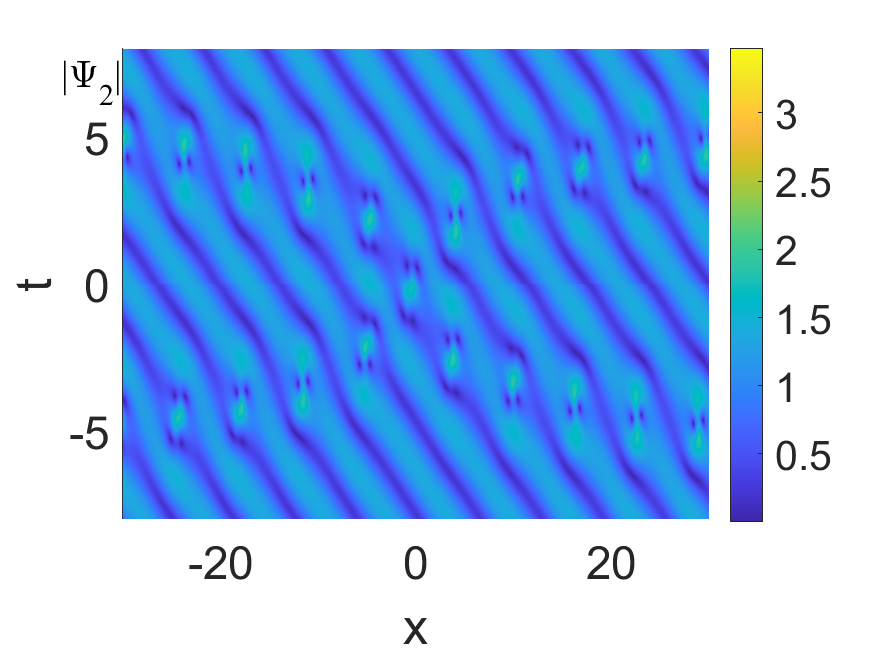}\newline
\vspace{-3mm}
\caption{The same as in Fig. \protect\ref{fig13}, but for the DP breathers
created on top of the space-only periodic background (top) and
spatiotemporal periodic background (bottom), whose unperturbed form is
displayed in Figs.~\protect\ref{fig9}(a) and \protect\ref{fig11}(a),
respectively.}
\label{fig14}
\end{figure}

The analytical results obtained above and stability of MP solitons/breathers
have been verified by numerical simulations, including added perturbations.
Employing the split-step Fourier method and fourth-order Runge-Kutta scheme~%
\cite{Yang}, initial conditions are set using the exact analytical solutions
obtained above at $t=0$ . The numerically produced evolution of
representative MP solitons/breathers, under the action of weak
perturbations, is shown in Figs.~\ref{fig12}-\ref{fig14}. It corroborates
excellent agreement of the numerical results with the analytical
predictions.

We have found that both the DP and triple-pole bright stripe solitons with the zero background,
as well as the MP beating stripe solitons with the nonzero plane backgrounds and MP breathers with
the spatially periodic background, exhibit robustness against perturbations. Accordingly, it is
demonstrated in Figs.~\ref{fig12}-\ref{fig14} that various MP solitons and breathers maintain
stable transmission under the action of initial random perturbation at the $2$\% level.

\section{Conclusions}

In this work, we have conducted a comprehensive investigation of the
dynamics of various MP (multi-pole) solitons and breathers modulated by the
helicoidal SOC (spin-orbit coupling in binary BECs (Bose-Einstein
condensates)) with attractive and repulsive inter-atomic interactions. Based
on the gauge transformation applied to the integrable Manakov system, we
have constructed exact analytical solutions for MP solitons and breathers,
followed by the rigorous analytical consideration of their dynamical
properties, in the cases of various backgrounds, \textit{viz}., MP stripe
solitons with zero background, MP beating stripe solitons on top of nonzero
plane-wave backgrounds, as well as MP beating stripe solitons and MP
breathers on spatiotemporal periodic backgrounds. The asymptotic analysis
reveals curved asymptotic trajectories for MP solitons/breathers (with the
additional straight trajectory of the central soliton pulse, in the case of
for odd-order poles), in distinct contrast with the straight trajectories of
conventional multi-solitons/breathers and periodic attraction-repulsion
trajectories of solitons/breathers which form bound states. The helicoidal
SOC induces spatially periodic stripe patterns in zero-background bright
solitons with the amplitude controlled at the origin, while enabling the
formation of solitons with the nonzero background, as the dark/bright
superpositions. These findings exhibit the coexistence of spatiotemporal
stripes and beating structures, whose periods are controlled by the SOC and
spectral parameters. Furthermore, wavenumber-matched/mismatched regimes
yield, respectively, beating stripe solitons and breathers with the periodic
background. Although helicoidal SOC generates diverse periodic out-of-phase
structures in the two BEC\ components, the total density remains nonperiodic
due to the mutual cancellation of the modulation in the components.
Numerical simulations have verified the analytical results and demonstrated
the stability of MP solitons/breathers against small perturbations.

It is essential to mention that, while this work addresses the solutions for
single MP solitons and breathers, the general MP solution, provided by
expression~(\ref{multi-pole}), allows straightforward derivation of $n$%
-solitons/breathers states of order $m_{k}+1$ for any $n\geq 1$ and $%
m_{k}\geq 1$. In particular, this framework
facilitates studies of interactions between MP solitons/breathers with
distinct orders.

From the experimental standpoint, the multi-pole solitons and breathers
reported here are accessible in spin-orbit-coupled Bose-Einstein condensates. The
helicoidal SOC can be realized via the spatially modulated Raman coupling
\cite{Lin2011,Kartashov2017}, and the excitation of these states is feasible
through phase-imprinting or density-engineering techniques, akin to those used for
generating dark-bright solitons. Their robustness against perturbations, as
confirmed by our numerics, and the tunability of their properties via SOC and
spectral parameters facilitate their experimental detection by means of the
time-of-flight or in-situ imaging observations.


\begin{acknowledgments}
This work was supported by the National Natural Science Foundation of China (Grant Nos. 12505006, 11975172, 12261131495 and 12381240286, the Israel Science Foundation (Grant No. 1695/22)), and by the Wuhan Textile University Special Fund Project.
\end{acknowledgments}

\vspace{5mm}\noindent{\large{\bf Declaration of competing interest}}\vspace{3mm}

The authors declare that they have no known competing financial interests or personal relationships that could have appeared to influence the work reported in this paper.

\appendix

\section{Lax pair for the Manakov system}

%\setcounter{equation}{0} \renewcommand\theequation{A\arabic{equation}}
%（A1）

The Lax pair of the Manakov system is taken as~\cite{Guo2012}
\begin{equation}
\begin{aligned}
\boldsymbol{\Phi_x}=&\boldsymbol{M\Phi},~~\boldsymbol{M}\equiv i(\lambda
\boldsymbol{\sigma}+\boldsymbol{U}),\\
\boldsymbol{\Phi_t}=&\boldsymbol{N\Phi},~~\boldsymbol{N}\equiv
i(\lambda^2\boldsymbol{\sigma}+\lambda
\boldsymbol{U})+\frac{1}{2}\boldsymbol{\sigma}(\boldsymbol{U_x}-i%
\boldsymbol{U}^2), \end{aligned}  \label{A1}
\end{equation}%
where
\begin{equation}
\begin{aligned} \boldsymbol{U}=\begin{pmatrix} 0 & su_1^* & su_2^* \\ u_1 &
0 & 0 \\ u_2 & 0 & 0
\end{pmatrix},~~\boldsymbol{\sigma}=\text{diag}(1,-1,-1), \end{aligned}
\end{equation}
and $\lambda$ represents the complex spectral parameter.

\section{Some parameters used in the main text}

Parameters in Eq.~(\ref{triple-bb-b3}) for the triple-pole stripe soliton
are given by
\begin{equation}
\begin{aligned} \label{parameter1}
\Gamma_1=&4\theta_R^4-8\theta_R^3+12\theta_R^2+128\lambda_I^4t^2(3-3%
\theta_R+\theta_R^2)+3,\\
\Gamma_2=&4\theta_R^4+8\theta_R^3+12\theta_R^2+128\lambda_I^4t^2(3-3%
\theta_R+\theta_R^2)+3,\\
\Delta_1=&2\theta_R^2-6\theta_R-16i\lambda_I^2t(\theta_R-2)+3,\\
\Delta_2=&2\theta_R^2+6\theta_R-16i\lambda_I^2t(-\theta_R-2)+3,\\
\Delta_3=&512\lambda_I^8t^4+2\theta_R^4+4\theta_R^2-32i\lambda_I^2t(%
\theta_R^2+1)\\ &+32\lambda_I^4t^2(2\theta_R^2+1)-3. \end{aligned}
\end{equation}%
Parameters in Eq.~(\ref{D-d-b}) for the DP beating stripe soliton are given
by
\begin{equation}
\begin{aligned} \label{parameter2}
\mathcal{D}_1=&2sz\lambda_I^2\delta^2-2\lambda_I\delta-2isz^3\lambda_I^2t+%
\lambda_I-z,\\
\mathcal{D}_2=&sz\lambda_I^2\delta^2-\lambda_I\delta+\lambda_I+(\beta-z)/2,%
\\ \mathcal{D}_3=&i(z\lambda_I^2\delta+s\beta). \end{aligned}
\end{equation}

\section{Double-pole dark $\Psi _{D}$ and bright $\Psi _{B}$ solitons in the
case of $|\protect\lambda _{I}|\leq 1$}

In this case, both DP dark $\Psi _{D}$ and bright $\Psi _{B}$ solitons in
solutions~(\ref{Double-pole dark-bright}) are generated with
parameterization $\lambda _{I}=\sin \gamma $ ($-\pi /2\leq \gamma \leq \pi /2
$). in the form of

\begin{equation}
\begin{aligned} \label{D-d-b2} \Psi_{D}=&e^{i\theta^\prime_1}-
\frac{8\lambda_I(2i\beta\Lambda^2e^{2\delta}+2i\lambda_I^4t^2+\mathcal{D}_1)
e^{i(\theta^\prime_1-\alpha)}} {\Lambda^{-2}\cos^2\gamma
e^{-2\delta}+4\Lambda^2e^{2\delta}+4(2\lambda_I^4t^2+\mathcal{D}_2)},\\
\Psi_{B}=&8\lambda_Ie^{i\theta^\prime_2}
\frac{2\Lambda(\lambda_I^2t+i\delta-i)e^\delta +\Lambda^{-1}(\lambda_I^2\beta
t-\mathcal{D}_3)e^{-\delta}} {\Lambda^{-2}\cos^2\gamma
e^{-2\delta}+4\Lambda^2e^{2\delta}+4(2\lambda_I^4t^2+\mathcal{D}_2)}
\end{aligned}
\end{equation}
where
\begin{equation}
\begin{aligned}
&\theta^\prime_1=\theta_1-k_mx,~~\mu_{1R}=-2\lambda_R-\cos\gamma\\
&\theta^\prime_2=\theta_1+k_mx+t/2+(-x+\mu_{1R}t)\cos\gamma,\\
&\delta=-\lambda_I(x-\mu_{1R}t),~~\beta=e^{-i\gamma}\cos\gamma,~~%
\Lambda=l_1/l_3,\\ &\mathcal{D}_1=2i\delta^2+2i(\beta-\lambda_I^2)\delta
+2\lambda_I^3(2ie^{-i\gamma}-\lambda_I)t+i\beta,\\
&\mathcal{D}_2=2\delta^2-2\lambda_I^2\delta+2\lambda_I^3t\cos\gamma+1,\\
&\lambda_I=\sin\gamma \end{aligned}
\end{equation}


\begin{thebibliography}{99}
\bibitem{Zakharov1972}
V. E. Zakharov, and A. B. Shabat, Exact theory of
two-dimensional self-focusing and one-dimensional selfmodulation of waves in
nonlinear media. Sov. Phys. JETP 34, 62-69 (1972).



\bibitem{SY}
J. Satsuma J. and N. Yajima, Initial value problems of
one-dimensional self-modulation of nonlinear waves in dispersive media,
Suppl. Prog. Theor. Phys. No. 55, 284-306 (1974).

\bibitem{Zakharov}
V. E. Zakharov, S. V. Manakov, S. P. Novikov, and L. P.
Pitaevskii, \textit{Theory of Solitons: The Inverse Problem Method }(Nauka
Publishers, Moscow, 1980) (English translation: Consultants Bureau, New
York, 1984).

\bibitem{Olmedilla1987}
E. Olmedilla, Multiple pole solutions of the
nonlinear Schr\"{o}dinger equation. Physica D 25, 330-346 (1987).

\bibitem{Gagnon1994}
L. Gagnon, and N. Stievenart, N-soliton interaction in
optical fibers: the multiple-pole case. Opt. Lett. 19, 619-621 (1994).

\bibitem{DP}
T. Dauxois and M. Peyrard, \textit{Physics of Solitons}
(Cambridge University Press, Cambridge, 2006).

\bibitem{Yang}
J. Yang, \textit{Nonlinear waves in integrable and nonintegrable systems} (SIAM, 2010).

\bibitem{Schiebold2017}
C. Schiebold, Asymptotics for the multiple pole
solutions of the nonlinear Schr\"{o}dinger equation. Nonlinearity, 30, 2930
(2017).

\bibitem{Zakharov2013} V. E. Zakharov, and S. Wabnitz, \textit{Optical
Solitons: Theoretical Challenges and Industrial Perspectives}
(Springer-Verlag, Heidelberg, 2013).

\bibitem{Kengne2023} E. Kengne, and W. Liu, \textit{Nonlinear waves: from
dissipative solitons to magnetic solitons} (Springer Nature, 2023).

\bibitem{Krolikowski1998} W. Kr\'{o}likowski, B. Luther-Davies, C. Denz, and
T. Tschudi. Annihilation of photorefractive solitons. Opt. Lett. 23, 97-99
(1998).

\bibitem{Poy2022} G. Poy, A. J. Hess, A. J. Seracuse, Mi. Paul, S. \v{Z}%
umer. and I. I. Smalyukh, Interaction and co-assembly of optical and
topological solitons. Nature Photon. 16, 454-461 (2022).

\bibitem{Zhang2024} M. Zhang, S. Ding, X. Li, K. Pu, S. Lei, M. Xiao,
and X. Jiang, Strong interactions
between solitons and background light in Brillouin-Kerr microcombs. Nature
Commun.\ 15, 1661 (2024).

\bibitem{Wang2025a}
D.L. Wang, Z.X. Liu,  H.P. Zhao, H.Q. Qin, C. He, C. Chen, P.J. Shi, W. Liu, D. Wang, G.Q. Zhou, X.M. He, C.Q. Dai, Launching by cavitation, Science 389(6763),  935-939 (2025)

\bibitem{Wang2025} Z. Wang, Y. Wang, B. Shi, C. Shen, W. Sun, Y. Ding, and
C. Bao, Rhythmic soliton interactions for integrated dual-microcomb
spectroscopy. Phys. Rev. X, 15(1), 011061 (2025).

\bibitem{Ablowitz2023} M. J. Ablowitz, J. T. Cole, G. A. El, M. A. Hoefer,
and X. D. Luo, Soliton-mean field interaction in Korteweg-de Vries
dispersive hydrodynamics. Stud. Appl. Math. 151(3), 795-856 (2023).

\bibitem{Abbagari2025} S. Abbagari, A. Houwe, L. Akinyemi, and S. Y. Doka,
Soliton interaction and nonlinear localized waves in one-dimensional
nonlinear acoustic metamaterials. Physica D, 476, 134591 (2025).

\bibitem{Ablowitz1992} M. J. Ablowitz, and P. A. Clarkson, \textit{Solitons,
Nonlinear Evolution Equations and Inverse Scattering} (Cambridge: Cambridge
University Press, 1992).

\bibitem{Zakharov1972a} V. E. Zakharov and A. B. Shabat, Interaction between
solitons in a stable medium. Sov. Phys. JETP, 64, 1627-1639 (1973).

\bibitem{Xin2022} F. Xin, L. Falsi, D. Pierangeli, F. Fusella, G.
Perepelitsa, Y. Garcia, and E. DelRe, Intense wave formation from multiple
soliton fusion and the role of extra dimensions. Phys. Rev. Lett. 129(4),
043901 (2022).

\bibitem{Qin2019} Y. H. Qin, L. C. Zhao, and L. Ling, Nondegenerate
bound-state solitons in multicomponent Bose-Einstein condensates. Phys. Rev.
E, 100(2), 022212 (2019).

\bibitem{Cui2024} Y. Cui, X. Yao, X. Hao, Q. Yang, D. Chen, Y. Zhang, and B.
A. Malomed, Dichromatic soliton-molecule compounds in mode-locked fiber
lasers. Laser Photonics Rev. 18(6), 2300471 (2024).

\bibitem{Li2020} M. Li, X. Yue, and T. Xu, Multi-pole solutions and their
asymptotic analysis of the focusing Ablowitz-Ladik equation. Phys. Scr.
95(5), 055222 (2020).

\bibitem{Li2020a} M. Li, X. Zhang, T. Xu, and L. Li, Asymptotic analysis and
soliton interactions of the multi-pole solutions in the Hirota equation. J.
Phys. Soc. Jpn. 89(5), 054004 (2020).

\bibitem{Xu2019a} T. Xu, Y. Chen, M. Li, and D. X. Meng, General stationary
solutions of the nonlocal nonlinear Schr\"{o}dinger equation and their
relevance to the PT-symmetric system. Chaos 29, 123124 (2019).

\bibitem{Poppe1983} C. Poppe, Construction of solutions of the sine-Gordon
equation by means of Fredholm determinants, Physica D 9, 103-139 (1983).

\bibitem{Wadati1981} M. Wadati, and K. Ohkuma, Multiple pole solutions of
the modified Korteweg-de Vries equation, J. Phys. Soc. Jpn. 51(6), 2029-2035
(1981).

\bibitem{Wang2024} S. F. Wang, Multipole solitons and vortex solitons in
nonlocal nonlinear media. Opt. Express, 32(9), 16132-16139 (2024).

\bibitem{Zhao2020} Z. Yan, and S. Y. Lou, Special types of solitons and
breather molecules for a (2+1)-dimensional fifth-order KdV equation, Commun.
Nonlinear Sci. Numer. Simul. 91, 105425 (2020).

\bibitem{Nguyen2019} T. V. Nguyen, Existence of multi-solitary waves with
logarithmic relative distances for the NLS equation. C. R. Math. 357, 13-58
(2019).

\bibitem{Rao2020} J. G. Rao, J. S. He, T. Kanna and D. Mihalache, Nonlocal $%
M $-component nonlinear Schr\"{o}dinger equations: Bright solitons,
energy-sharing collisions, and positons. Phys. Rev. E 102, 032201 (2020).

\bibitem{Rao2021} J. G. Rao, T. Kanna, K. Sakkaravarthi and J. S. He,
Multiple double-pole bright-bright and bright-dark solitons and
energy-exchanging collision in the $M$-component nonlinear Schr\"{o}dinger
equations. Phys. Rev. E 103, 062214 (2021).

\bibitem{Gordon1983} J.P. Gordon, Interaction forces among solitons in
optical fibers. Opt. Lett. 8, 596-598 (1983)

\bibitem{Karlsson1996} M. Karlsson, D. J. Kaup, B.A. Malomed, Interactions
between polarized soliton pulses in optical fibers: exact solutions. Phys.
Rev. E 54, 5802-5808 (1996)

\bibitem{Xu2019} G. Xu, A. Gelash, A. Chabchoub, V. Zakharov, and B. Kibler,
Breather wave molecules. Phys. Rev. Lett. 122(8), 084101 (2019).

\bibitem{Kedziora2012} D. J. Kedziora, A. Ankiewicz, and N. Akhmediev,
Second-order nonlinear Schr\"{o}dinger equation breather solutions in the
degenerate and rogue wave limits, Phys. Rev. E 85, 066601 (2012).

\bibitem{Liu2017} T. Y. Liu, T. L. Chiu, P. A. Clarkson, and K. W. Chow, A
connection between the maximum displacements of rogue waves and the dynamics
of poles in the complex plane. Chaos, 27(9), 091103 (2017).

\bibitem{Lin2011} Y. J. Lin, K. Jim\'{e}nez-Garc\'{\i}a, and I. B. Spielman,
Spin-orbit-coupled Bose-Einstein condensates, Nature (London), 471, 83
(2011).

\bibitem{Zhai2015} H. Zhai, Degenerate quantum gases with spin-orbit
coupling: A review, Rep. Prog. Phys. 78, 026001 (2015).

\bibitem{Zhang2016} Y. Zhang, M. E. Mossman, T. Busch, P. Engels, and C.
Zhang, Properties of spin-orbit-coupled Bose-Einstein condensates, Front.
Phys. 11, 118103 (2016).

\bibitem{Dalibard2011} J. Dalibard, F. Gerbier, G. Juzeliunas, and P. \"{O}%
hberg, Colloquium: Artificial gauge potentials for neutral atoms, Rev. Mod.
Phys. 83, 1523 (2011).

\bibitem{Ruseckas2005} J. Ruseckas, G. Juzeliunas, P. \"{O}hberg, and M.
Fleischhauer, Non-Abelian gauge potentials for ultracold atoms with
degenerate dark states, Phys. Rev. Lett. 95, 010404 (2005).

\bibitem{Kartashov2017} Y. V. Kartashov and V. V. Konotop, Solitons in
Bose-Einstein condensates with helicoidal spin-orbit coupling, Phys. Rev.
Lett. 118, 190401 (2017).

\bibitem{Kartashov2019} Y. V. Kartashov, V. V. Konotop, M. Modugno, and E.
Ya. Sherman, Solitons in inhomogeneous gauge potentials: Integrable and
nonintegrable dynamics, Phys. Rev. Lett. 122, 064101 (2019).

\bibitem{Fang2024} P. Fang and J. Lin, Soliton in Bose-Einstein condensates
with helicoidal spin-orbit coupling under a Zeeman lattice, Phys. Rev. E
109, 064219 (2024).

\bibitem{Yang2022}
Y. Yang, P. Gao, L.-C. Zhao, and Z.-Y. Yang, Kink-like breathers in Bose-Einstein condensates with helicoidal spin-orbit coupling, Front. Phys. 17, 32503 (2022).



\bibitem{Jimenez2015} K. Jim\'{e}nez-Garc\'{\i}a, L. J. LeBlanc, R. A.
Williams, M. C. Beeler, C. Qu, M. Gong, C. Zhang, and I. B. Spielman,
Tunable spin-orbit coupling via strong driving in ultracold-atom systems,
Phys. Rev. Lett. 114, 125301 (2015).

\bibitem{Luo2016} X. Luo, L. Wu, J. Chen, Q. Guan, K. Gao, Z.-F. Xu, L. You,
and R. Wang, Tunable atomic spin-orbit coupling synthesized with a
modulating gradient magnetic field, Sci. Rep. 6, 18983 (2016).

\bibitem{Samsonov2004} S. V. Samsonov, A. D. R. Phelps, V. L. Bratman, G.
Burt, G. G. Denisov, A. W. Cross, K. Ronald, W. He, and H. Yin, Compression
of frequency-modulated pulses using helically corrugated waveguides and its
potential for generating multigigawatt rf radiation, Phys. Rev. Lett. 92,
118301 (2004).

\bibitem{Burt2004} G. Burt, S. V. Samsonov, K. Ronald, G. G. Denisov, A. R.
Young, V. L. Bratman, A. D. R. Phelps, A. W. Cross, I. V. Konoplev, W. He,
J. Thomson, and C. G. Whyte, Dispersion of helically corrugated waveguides:
Analytical, numerical, and experimental study, Phys. Rev. E 70, 046402
(2004).

\bibitem{Manakov1974} S. V. Manakov, On the theory of two-dimensional
stationary self-focusing of electromagnetic waves, Sov. Phys. JETP 38, 248
(1974).

\bibitem{Kartashov2014}
Y. V. Kartashov, V. V. Konotop, and D. A. Zezyulin, Bose-Einstein condensates with localized spin-orbit coupling: Soliton complexes and spinor dynamics, Phys. Rev. A 90, 063621 (2014).



\bibitem{Guo2012} B. Guo, L. Ling, and Q. P. Liu, Nonlinear Schr\"{o}dinger
equation: generalized Darboux transformation and rogue wave solutions, Phys.
Rev. E 85, 026607 (2012).

\bibitem{ding2025} C. C. Ding, Q. Zhou, and B. A. Malomed, Beating stripe
solitons arising from helicoidal spin-orbit coupling in Bose-Einstein
condensates, Phys. Rev. E 111, 044203 (2025).
\end{thebibliography}
\end{document}